\documentclass[apj,chicago]{emulateapj}
\usepackage{times,mathptmx}

\shorttitle{{\sc Protoplanets in turbulent isothermal disks}}
\shortauthors{{\sc Baruteau \& Lin}}

\begin{document}
\title{Protoplanetary migration in turbulent isothermal disks}
\author{C. Baruteau\altaffilmark{1} and D.N.C. Lin\altaffilmark{1,2}}
\affil{$^1$Astronomy and Astrophysics Department, University of California, Santa Cruz, CA 95064, USA\\
$^2$Kavli Institute for Astronomy and Astrophysics, Peking University, Beijing, China}
\email{clement.baruteau@ucolick.org; lin@ucolick.org}

\keywords{accretion, accretion disks --- hydrodynamics --- turbulence --- methods: numerical --- planetary systems: formation --- planetary systems: protoplanetary disks}

\begin{abstract}
In order to reproduce the statistical properties of the observed exoplanets, population synthesis models have shown that the migration of protoplanets should be significantly slowed down, and that processes stalling migration should be at work. Much current theoretical efforts have thus been dedicated to find physical effects that slow down, halt or even reverse migration. Many of these studies rely on the horseshoe drag, whose long term-evolution (saturated or not) is intimately related to the disk viscosity in laminar disk models. We investigate how the horseshoe drag exerted on a low-mass planet is altered by a more realistic treatment of the turbulence in protoplanetary disks. Two-dimensional hydrodynamic simulations are performed with a turbulence model that reproduces the main turbulence properties of three-dimensional magnetohydrodynamic calculations. We find that the horseshoe drag can remain unsaturated on the long term, depending on the turbulence strength. We show that the desaturation of the horseshoe drag by turbulence can be modeled by vortensity diffusion across the time-averaged planet's horseshoe region. At low-turbulence, the running-time averaged torque is in good agreement with the total torque obtained for an equivalent laminar model, with a similar vortensity's diffusion coefficient. At high-turbulence, differences arise due to the time-evolution of the averaged density profile with turbulence.
\end{abstract}

\section{Introduction}
The discovery of exoplanets has provided an exciting opportunity to test our theories of planet formation and evolution. The detection of the first hot Jupiter by \cite{MQ95} made planetary migration become a central ingredient of these theories \citep{lin96}. Planetary migration occurs from the early stages of planet formation, as planets tidally interact with the protoplanetary disk where they are embedded in \citep{gt80}. Planets experience a torque from the disk that alters their eccentricity and semi-major axis. The variation of the planets semi-major axis under the disk torque is referred to as planetary migration.

Depending on the disk and planet properties, three types of migration are usually distinguished. Type I migration applies to low-mass planets, typically up to a few Earth masses if the central object has a solar mass. In isothermal disks, type I migration is directed inwards and is much faster than both the formation of giant planets cores and the disk evaporation \citep[e.g.][]{w97, tanaka2002}. To reproduce the mass-distance distribution of the observed exoplanets, population synthesis models have shown that type I migration should be significantly slowed down \citep{IdaLin4, Mordasini09b}, and that processes stalling migration should be at work, in particular near the snow line \citep{IdaLin5, sli09}. Some mechanisms recently proposed to slow down, halt or even reverse migration are reviewed below. Type II migration concerns planets that are massive enough to open up a clean gap around their orbit (typically Jupiter-mass planets). In this regime, planets drift inwards at the disk viscous rate \citep{lp86}. Intermediate-mass planets building up a partial gap can undergo a runaway migration in massive disks \citep{mp03}. This is usually referred to as type III migration. Throughout this paper, we will focus on type I migration.

\subsection{Type I migration}
\label{sec:typeImig}
The torque exerted on a low-mass planet comprises the differential Lindblad torque and the horseshoe drag \citep[e.g.][]{pp09a}. The former results from the exchange of angular momentum between the planet and the circulating\footnote{In the frame corotating with the planet.} fluid elements located at Lindblad resonances \citep{gt79}. The differential Lindblad torque is generally negative \citep{w97}, and is stationary after a few dynamical timescales \citep{mvs87}. Its value is however particularly sensitive to any mechanism altering the location of the resonances, such as the disk self-gravity \citep{ph05, bm08b}. 

The horseshoe drag accounts for the exchange of angular momentum between the planet and the librating\footnotemark[1] fluid elements in the planet's horseshoe region \citep{wlpi91}. \cite{cm09} have clarified that the horseshoe drag actually corresponds to the corotation torque, namely the torque exerted by the corotation region\footnote{As pointed out by \cite{cm09}, the corotation and horseshoe regions generally differ. While a linear analysis indicates that the corotation region has a radial width of approximately a pressure scaleheight $H$ about the corotation radius, the width of the horseshoe region vanishes as the planet's mass tends to zero.} on the planet. The horseshoe drag is a non-linear phenomenon that cannot be captured by linear theory \citep{pp09a}, except at early times, before fluid streamlines get fully adjusted to the planet's introduction in a timescale comparable to the horseshoe U-turn time \citep{bm08a}. Contrary to the differential Lindblad torque, the horseshoe drag can be either positive or negative, and it is generally not a stationary quantity. In isothermal disks, its time dependence is related to the time evolution of the disk vortensity (the vorticity to surface density ratio) in the horseshoe region. The maximum value of the horseshoe drag scales with the (opposite of the) unperturbed vortensity gradient across the horseshoe region \citep{wlpi92, masset01}. This maximum value is called the fully unsaturated horseshoe drag. It is positive for surface density profiles locally shallower than $r^{-3/2}$. In the particular case of a planet fixed in an inviscid disk, the horseshoe region is closed and vortensity is conserved along horseshoe streamlines. As vortensity is progressively stirred up in the horseshoe region, the horseshoe drag oscillates with time with a decreasing amplitude. The horseshoe drag ultimately cancels out as vortensity gets uniformly distributed after several libration times \citep{bk2001}. This is known as the horseshoe drag saturation. 

For radiative disks in the adiabatic limit, the horseshoe drag has an additional contribution, whose maximum value scales with the (opposite of the) unperturbed entropy gradient across the horseshoe region \citep{bm08a, pm08, mc09}. This additional contribution is called entropy-related horseshoe drag. The contribution already existing in isothermal disks, and which features the vortensity gradient, is similarly denoted as vortensity-related horseshoe drag for future reference. The entropy-related horseshoe drag is positive for negative entropy gradients. It can therefore slow down, halt or even reverse \citep{pm06} type I migration compared to isothermal disk models, as long as saturation does not occur.

\subsection{Horseshoe drag desaturation}
\label{sec:hdd}
As described above, the horseshoe drag in isothermal disks saturates as vortensity is strictly advected along horseshoe streamlines. Viscosity, which acts as a source term in the vortensity equation \citep[e.g.][]{og2006}, can sustain a non-zero vortensity gradient across the horseshoe region. If so, the vortensity-related horseshoe drag would attain a steady-state value \citep{masset01, masset02}, which arises from a net exchange of angular momentum between the horseshoe region and the rest of the disk. In particular, the torque remains unsaturated when the viscous diffusion time across the horseshoe region is somewhat larger than the horseshoe U-turn time \citep{bm08a}, but smaller than the libration time \citep[e.g.][]{wlpi91, masset01}. At large viscosities, the horseshoe drag decreases with increasing viscosity, as viscous diffusion tends to impose the initial vortensity profile. This behavior is known as the cut-off of the horseshoe drag \citep{masset01}. For strong enough viscosities, the horseshoe drag coincides with the linear estimate of the corotation torque \citep{pp09a}.

It is still quite uncertain how the horseshoe drag desaturation operates in radiative disks. The vortensity-related horseshoe drag presumably requires the same constraint on viscosity as in isothermal disk models. Similarly, \cite{bm08a} argued that the entropy-related horseshoe drag should remain fully unsaturated if the timescale for restoring the unperturbed entropy gradient is also bound by the horseshoe U-turn and libration timescales. This result has been independently verified by \cite{pp08} and by  \cite{KleyCrida08}, who respectively considered thermal diffusion and radiative effects as the dominant physical process sustaining the entropy gradient. However, the desaturations of both parts of the horseshoe drag are likely to be coupled. In particular, some viscosity should always be required to unsaturate the entropy-related horseshoe drag, so as to allow the exchange of angular momentum between the horseshoe region and the rest of the disk. A systematic study involving different viscous and thermal diffusion coefficients will certainly give more insight into the desaturation process in radiative disks.

The horseshoe drag also plays a key role in another mechanism potentially halting migration. \cite{masset06a} showed that type I migration can be stalled near positive surface density transitions, when planets migrate toward underdense regions in the protoplanetary disk. These density transitions can be caused by ionization transitions, which may occur near the snow line \citep{kl07}, or near the inner edge of the dead zone (the region near the disk mid-plane sandwiched together by partially ionized surface layers). As the planet drifts inwards, the slope of the local density profile becomes increasingly positive. The positive horseshoe drag strongly increases, and migration halts when it balances the negative differential Lindblad torque. The ability of these so-called planet traps to stall migration depends on the horseshoe drag amplitude, hence on the local dissipation processes regulating its desaturation.

\subsection{Turbulence}
Whether one invokes the gas thermodynamics to slow down, halt or reverse migration, or the presence of planet traps to stall migration, dissipation processes, including viscosity, must be taken into account. In the framework of planet migration, protoplanetary disks are commonly assumed to be laminar, with a constant viscosity whose amplitude aims at modeling the turbulent transport of angular momentum.

Turbulence may originate from hydrodynamic instabilities, such as the Rossby-wave instability \citep{lovelace99}, the global baroclinic instability \citep{kb2003}, or the Kelvin-Helmholtz instability, triggered by the vertical shear of the gas as dust settles in the mid-plane \citep{jhk06}. Another source of turbulence is the so-called magnetorotational instability \citep[MRI,][]{bh91}. This instability relies on the coupling of the ionized gas to the weak magnetic field of the disk. Ionization may occur in the vicinity of the central object due to the star irradiation, or further out in the disk layers, most probably through the UV background or cosmic rays. The ionization state of the disk mid-plane, where protoplanets build up and migrate, is rather unclear. It is still a matter of debate whether the disk mid-plane should be ionized as well, or if it remains neutral, being shielded by the layers. In any case, its ionization state may significantly evolve with time \citep{IlgnerNelson08}. 

So far, only magnetohydrodynamic (MHD) turbulence has been extensively studied in numerical simulations of planet-disk interactions. \cite{qmwmhd2}, \cite{wbh03} and \cite{pap2004} have investigated the properties of gap formation by giant planets in turbulent disks. \cite{qmwmhd4} and \cite{nelson05} have revisited type I migration for one or several planets embedded in turbulent disks. They found that low-mass planets evolve on a random walk, in contrast to the systematic decay expected in laminar disks. It is unclear whether, with longer simulations, the time-averaged stochastic torque due to turbulence would become negligible compared to the total mean torque. Another open question is whether this total mean torque significantly differs from that obtained in laminar disks. In this regard, a systematic study of the horseshoe drag behavior with turbulence has not been done yet. We eventually point out that all aforementioned studies have assumed a fully magnetized disk. Simulations of planets interacting with disks harboring a dead zone have not been performed yet.

\subsection{Paper outline}
In this paper, we investigate the impact of turbulence on the horseshoe drag. Can the horseshoe drag remain unsaturared with turbulence? Can migration be stalled at planet traps with turbulence? Three-dimensional MHD calculations would certainly be an elegant way to answer these questions. Nonetheless, the horseshoe drag saturation typically occurs on several hundreds of orbits for planets subject to type I migration. The computational cost of 3D MHD calculations properly resolving the horseshoe region over several hundreds of orbits is prohibitive. We perform instead two-dimensional hydrodynamic simulations using the model of \cite{lsa04} (hereafter LSA04) for mimicking the properties of MHD turbulence. Further simplification is obtained by assuming an isothermal disk. The inclusion of the disk thermodynamics will be investigated in a subsequent study.

The plan of the paper is as follows. In \S~\ref{sec:setup}, we describe the numerical set-up of our simulations. The turbulence model used in our study is detailed in \S~\ref{sec:turb}, in particular we assess the minimum duration of the calculations to reach torque convergence. Our calculation results on the horseshoe drag desaturation with turbulence are presented in \S~\ref{sec:numres}. Discussion and conclusions follow in \S~\ref{sec:discuss}.

\section{Numerical set-up and notations}
\label{sec:setup}
To investigate the desaturation properties of the horseshoe drag with turbulence, we perform two-dimensional hydrodynamic simulations using the turbulence model of LSA04. Before detailing the physical properties of this potential in \S~\ref{sec:turb}, we describe in this section the numerical set-up of our calculations.

We use the two-dimensional code FARGO\footnote{See: {\texttt http://fargo.in2p3.fr}} in its isothermal version. It is a staggered mesh code that solves the Navier-Stokes and continuity equations on a polar grid. An upwind transport scheme is used along with a harmonic, second-order slope limiter \citep{vl77}. Its specificity is to use a change of rotating frame on each ring of the grid, which increases the timestep significantly \citep{fargo1,fargo2}. 

The code units are the following. The initial orbital radius $r_p$ of the planet is the length unit, the mass of the central object $M_*$ is the mass unit, and $(GM_*/ {r_p}^3)^{-1/2}$ is the time unit, $G$ being the gravitational constant ($G = 1$ in our unit system). The orbital period at $r=r_p$ is denoted by $T_{\rm orb}$.

We denote the polar coordinates by $r$ and $\varphi$. The disk is initially axisymmetric and rotates at the sub-Keplerian angular velocity $\Omega(r)$. The disk is isothermal, its aspect ratio reads $h(r) = h_p \times (r / r_p)^{1/2}$, with $h_p$ ranging from $3\%$ to $7\%$. The initial surface density is $\Sigma_p \times(r/r_p)^{-\sigma}$, with $\Sigma_p = 5\times 10^{-4}$ and $\sigma = 0.5$ by default. The smallest value of the Toomre parameter at $r=r_p$ then amounts to $Q_p \sim 20$. The disk self-gravity is thus neglected throughout this study. Should we have included the full self-gravity, the differential Lindblad torque would have been larger by less than $5\%$ \citep[][figure 8]{bm08b}. No kinematic viscosity is included when the turbulent potential is applied to the disk. For comparison, runs without turbulence, but with a constant kinematic viscosity $\nu$ are also performed. They will be referred to as laminar runs, whereas calculations featuring the turbulent potential will be mentioned as turbulent runs for future reference. 

The planet mass is denoted by $M_p$, and $q$ is the planet to primary mass ratio. The planet's Bondi radius $r_B$ is defined as $r_B = GM_p / c_s^2$, where $c_s$ is the sound speed at the planet location. The softening length of the planet's potential is $\varepsilon = 0.6h_pr_p$, so that $r_B / \varepsilon \approx 1.7 (q / h_p^3)$. For the planet masses considered in \S~\ref{sec:numres}, $r_B / \varepsilon$ ranges from 0.3 to 0.6.

The grid used in our calculations has $N_r = 512$ radial zones, and $N_s = 1536$ azimuthal zones. The disk extends from $r_{\rm min} = 0.4\,r_p$ to $r_{\rm max} = 1.8\,r_p$ along the radial direction, and wave-killing zones are used next to the boundaries to minimize unphysical wave reflexions \citep{valborro06}.

\section{Turbulence properties}
\label{sec:turb}
The turbulence model used in our study is that of LSA04. It is based on applying an external {\it turbulent} potential to the disk. We show in this section that the properties of this potential (spatial and temporal fluctuations, amplitude) can be tuned so that the perturbations it induces much resemble those obtained with 3D MHD calculations. As in the original work of LSA04, the turbulent potential $\Phi_{\rm turb}$ applied to the disk corresponds to the superposition of 50 simultaneous wave-like modes. It reads \citep[][hereafter OIM07]{oim07}:
\begin{equation}
	\Phi_{\rm turb}(r,\varphi,t) = \gamma r^2 \Omega^2 \sum_{k=1}^{50} \Lambda_{k}(r,\varphi,t),
\label{turbpot}
\end{equation}
where $\gamma$ is a dimensionless constant indicating the turbulence strength, and
\begin{equation}
\Lambda_{k} = \xi_k\,e^{-(r-r_{k})^2 / \sigma_k^2}\cos(m(k)\varphi - \varphi_{k} - \Omega_{k}\tilde{t_k})\sin(\pi\tilde{t_k} / \Delta t_k).
	\label{lambda}
\end{equation}
In Equation~(\ref{lambda}), $(r_k, \varphi_{k})$ denotes the initial location of the mode of wavenumber $m(k)$. Both $r_k$ and $\varphi_{k}$ are randomly sorted with a uniform distribution. The modes radial extent is $\sigma_k = \pi r_k / 4m$. Modes start at time $t_{0,k}$, and their lifetime is $\Delta t_k = 2\pi r_k / m c_s$, with $c_s$ the local sound speed. We denote by $\Omega_k$ the Keplerian angular frequency at $r=r_k$, $\tilde{t_k} = t - t_{0,k}$, and $\xi_k$ is a dimensionless constant sorted randomly with a Gaussian distribution of unit width. Note that the expression of the turbulent potential is independent of the disk surface density.

\subsection{Power spectrum and modes truncation}
\begin{figure}
  \includegraphics[width=\hsize]{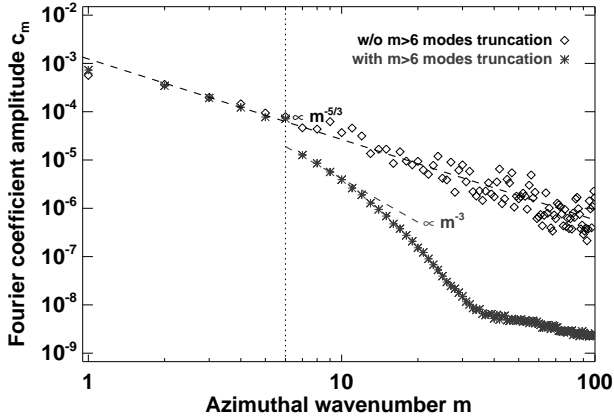}
  \caption{\label{spectrum}Time-averaged power spectrum of the density perturbations due to turbulence. Results are shown with the $m>6$ modes truncation (stars), and without it (diamonds). Both calculations have $\gamma = 4\times 10^{-5}$. No planet is included.}
\end{figure}

LSA04 randomly sorted $m$ with a logarithmic distribution between $m=2$ and $m=N_{\sec}/8$. Following OIM07, we also include $m=1$ modes, and we set $\Lambda_{k} = 0$ if $m(k) > 6$. The latter assumption is motivated by the fact that high-$m$ modes contribute less to the turbulent potential, since $\Lambda_{k} \propto \exp(-m^2)$, within smaller time intervals ($\Delta t_k \propto 1/m$). They therefore contribute less to the turbulent torque exerted on the planet. As we shall see hereafter, excluding $m>6$ modes helped reduce the convergence time of our calculations. This cut-off impacts the power spectrum of the density perturbations triggered by the turbulent potential. The amplitude $c_m$ of the corresponding Fourier coefficients, which read
\begin{equation}
c_m(t) = \frac{ | \int_{r_{\rm min}}^{r_{\rm max}}\int_0^{2\pi} \Sigma(r,\varphi,t)\,rdr\,d\varphi\,e^{-im\varphi} | }
{ \int_{r_{\rm min}}^{r_{\rm max}}\int_0^{2\pi} \Sigma(r,\varphi,t)\,rdr\,d\varphi },
\label{eqn:cm}
\end{equation}
were calculated for two simulations with $\gamma = 4\times 10^{-5}$, and with no planet. One includes the $m>6$ modes truncation, the other does not. We display in Figure~\ref{spectrum} the coefficients $c_m$ time-averaged over the runs duration (about $700\,T_{\rm orb}$). The density spectrum without truncation, and that of the 3D MHD simulation of LSA04 (see their figure 2), have very similar slopes. The upper dashed line overplotted in Figure~\ref{spectrum} indicates that this spectrum decreases as $m^{-5/3}$. We comment that in absence of modes cut-off in the turbulent potential, the power spectrum features both the modes that are directly forced by the turbulent potential, and the modes induced by energy cascade from larger scales. A detailed comparison with Kolmogorov theory of weakly compressible fluids turbulence would not be strictly relevant here. Spectra with and without the $m>6$ modes cut-off much resemble up to $m=6$, below which the spectrum with truncation drops off, and starts decreasing as $m^{-3}$. This is the slope expected in two-dimensional forward decaying turbulence \citep{k67}. The turning point from $m\approx 30$ is not expected, however. With increasing the turbulence amplitude to $\gamma = 10^{-4}$, we find that the turning point moves to a higher value ($m\approx 60$), while the spectrum without truncation is not significantly altered. Additionally, we checked that the spectrum of velocity perturbations is very similar to the density spectrum, as expected with an disk. Unless otherwise stated, all calculation results shown below are obtained with the $m > 6$ modes cut-off.

\subsection{Autocorrelation timescale}
\label{sec:act}
\begin{figure}
  \includegraphics[width=\hsize]{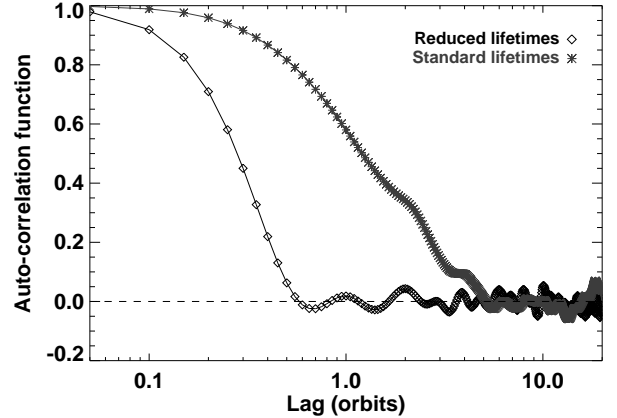}
  \caption{\label{autocor}Autocorrelation function of the torque per unit mass exerted by the turbulent disk. We compare the autocorrelation timescales obtained with the modes standard lifetimes (LSA04, stars) and reduced lifetimes (standard lifetimes divided by ten, diamonds).}
\end{figure}

The autocorrelation timescale $\tau_c$ measures the typical timescale over which perturbations due to turbulence become uncorrelated. Following \cite{oishi07}, we evaluate the autocorrelation function of the torque per unit mass, $\Gamma$, exerted by the turbulent disk on a massless planet. Its normalized value is given by
\begin{equation}
{\rm ACF} (\tau) = \frac{\int_{\tau}^{t_{\rm max}} \Gamma(t)\,\Gamma(t-\tau)\,dt}{\int_{\tau}^{t_{\rm max}} \Gamma^2(t)\,dt},
\label{eqn:acf}
\end{equation}
where $t_{\rm max}$ is the runs duration. The quantity ACF is displayed in Figure~\ref{autocor} for two calculations, one with the modes lifetimes taken by LSA04 ($\Delta t_k = 2\pi r_k / m c_s$, they are referred to as standard lifetimes), and one for which $\Delta t_k$ is reduced by a factor of ten (reduces lifetimes). For both runs, $\gamma = 4\times 10^{-5}$, and the disk aspect ratio at $r=r_p$ (planet's location if any) is $h_p = 3\%$. As in all other calculations, the sampling time of the torque is $T_{\rm orb}/20$. 

There are several methods for evaluating $\tau_c$. One is based on calculating $\tau_c$ as $\tau_c = \int_{0}^{t_{\rm max}}{\rm ACF}(\tau)d\tau$ \citep[e.g.][]{rp09}. With the standard lifetimes (denoted by stars in Figure~\ref{autocor}), we find $\tau_c \approx 2\,T_{\rm orb}$, whereas $\tau_c \approx 0.3\,T_{\rm orb}$ with the reduced lifetimes (diamonds in the same figure). Alternatively, $\tau_c$ can be estimated as the smallest lag value for which the autocorrelation function crosses zero with a positive slope (second zero-crossing, see e.g. \cite{oishi07}). From Figure~\ref{autocor}, this method yields $\tau_c \approx 6\,T_{\rm orb}$ with the standard lifetimes, and $\tau_c \approx 0.8\,T_{\rm orb}$ with the reduced lifetimes. The autocorrelation timescale obtained with the reduced lifetimes is in good agreement with the typical values obtained with 3D MHD calculations, be they local \citep{FroPap06, oishi07} or global \citep{qmwmhd4, fn09}, and be they obtained with disks fully invaded by the MRI, or including a dead zone \citep{oishi07}. Unless otherwise stated, the runs presented hereafter use the modes reduced lifetimes: $\Delta t_k = 0.2\pi r_k / m c_s$.

\subsection{Reynolds stress parameter}
\label{sec:alpha}
\begin{figure}
   \includegraphics[width=\hsize]{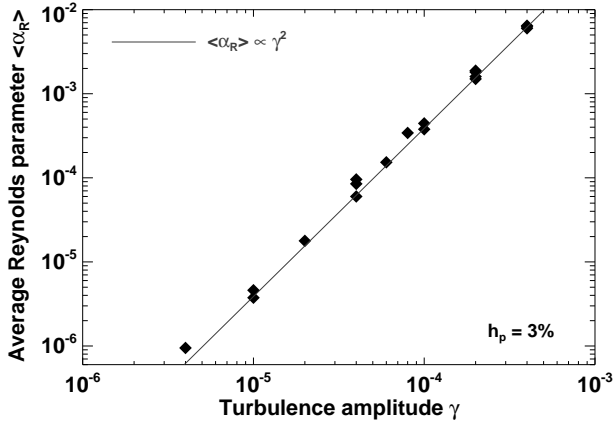}
   \caption{\label{viscequiv}Average Reynolds stress parameter $\langle\alpha_R\rangle$ calculated for several values of the turbulent coefficient $\gamma$. For all runs, the disk aspect ratio at $r=r_p$ is $h_p = 3\%$. The solid curve depicts a quadratic function of $\gamma$ passing through the point $(\gamma = 4\times 10^{-4}, \langle\alpha_R\rangle = 6.2\times 10^{-3})$.}
\end{figure}

Another relevant property of the turbulent potential is the transport of angular momentum it generates. The turbulence amplitude $\gamma$ can thus be related to the magnitude of $\alpha$ in the widely adopted alpha prescription for turbulent viscosity \citep{ss73}. A usual procedure is to calculate the Reynolds stress parameter $\alpha_R (r,t)$, which we evaluate as
\begin{equation}
	\alpha_R (r,t) = \frac{\tilde\Sigma\,\overline{\delta v_r \delta v_{\varphi}}}{\overline{P}},
\label{alphameasure}
\end{equation}
where $\tilde\Sigma$ denotes the axisymmetric surface density, $P = \Sigma c_s^2$ is the gas pressure, $\delta v_r = v_r - \overline{v_r}$ and $\delta v_{\varphi} = v_{\varphi} - \overline{v_{\varphi}}$, with $v_r$ ($v_{\varphi}$) the radial (azimuthal) component of the gas velocity. Here the overbar symbols denote surface density-weighted azimuthal averages. The numerator in Equation~(\ref{alphameasure}) features the Reynolds stress tensor. We furthermore average the profile $\alpha_R (r,t)$ (i) with radius over the whole disk, then (ii) with time over the calculations duration. The resulting number is denoted by $\langle\alpha_R\rangle$. We performed a series of runs with varying $\gamma$, following the set-up described in \S~\ref{sec:setup} ($h_p = 3\%$). We display $\langle\alpha_R\rangle$ as a function of $\gamma$ in Figure~\ref{viscequiv}. The solid curve highlights that $\langle\alpha_R\rangle$ scales as $\gamma^2$. This scaling is expected as both the perturbed velocities $\delta v_r$ and $\delta v_{\varphi}$ scale with $\gamma$. 

We investigated the dependence of $\langle\alpha_R\rangle$ with various parameters of the turbulent potential. Firstly, we found that, for a fixed value of $\gamma$, $\langle\alpha_R\rangle$ is typically increased by a factor $\sim 1.5$ without the $m>6$ modes truncation. This confirms that most of the angular momentum exchange driven by turbulence occurs through (very) low wavenumbers. Secondly, we varied $\tau_c$ by setting the modes lifetimes to $\Delta t_k = n\times 2\pi r_k / m c_s$, with $n$ varying from $0.02$ to $2$. By taking the second zero-crossing of the torque autocorrelation function, we measured $\tau_c$ in the range $[0.6-8]\,T_{\rm orb}$. Note that $\tau_c$ does not scale with $n$. We found that $\langle\alpha_R\rangle$ scales with $\tau_c$ for $\tau_c \lesssim T_{\rm orb}$. Beyond, $\langle\alpha_R\rangle$ slightly decreases with $\tau_c$ (by $\sim 20\%$ when increasing $\tau_c$ from $\approx 2\,T_{\rm orb}$ to $\approx 8\,T_{\rm orb}$). Lastly, another series of runs with several values of $\gamma$ was performed for $h_p = 7\%$. Altering $h_p$ (or, equivalently, the sound speed) modifies the pressure ($P \propto h_p^2$) and the modes lifetimes ($\Delta t_k \propto h_p^{-1}$). From the comparison with the previous series for $h_p = 3\%$, we found that $\langle\alpha_R\rangle$ scales with $h_p^{-2}$, suggesting the decrease of the modes lifetimes was here of negligible importance. We checked indeed that reducing $h_p$ from $7\%$ to $3\%$ led to decrease $\tau_c$ by only a $\sim 20\%$ factor. The results obtained with both series of runs therefore lead to the following relationship:
\begin{equation}
	\langle\alpha_R\rangle \approx 35 \left( \frac{\gamma}{h_p} \right)^2,
\label{alpha}
\end{equation}
which inverts as $\gamma \approx 1.7\times10^{-1}\,h_p\,\sqrt{\langle\alpha_R\rangle}$.

\subsection{Diffusion coefficient}
\label{sec:schmidt}
\begin{figure}
   \includegraphics[width=\hsize]{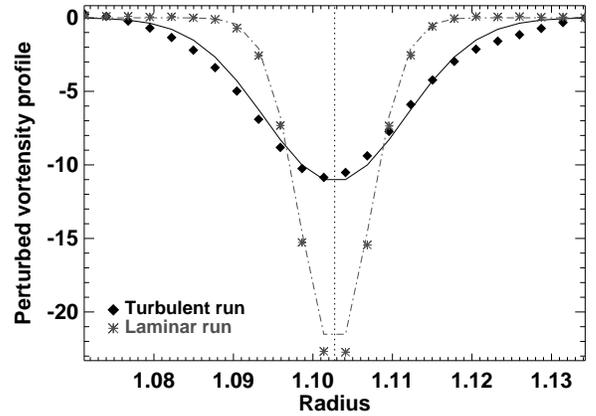}
   \caption{\label{schmidt}Vortensity profiles obtained for laminar and turbulent runs with the same equivalent Reynolds alpha parameter $\langle \alpha_R \rangle$. Both profiles are subtracted from the background profile. A slight negative perturbation ($10\%$ relative difference) is initially imposed to the unperturbed vortensity profile at $r=r_D$, the location of which is depicted by the vertical dotted line. Stars show the perturbed vortensity profile at $t=4.5\,T_{\rm orb}$ for a laminar run with $\nu = 3.4\times 10^{-7}$. The dash-dotted curve shows the profile expected from Equation~(\ref{vortdirac}) for $D = \nu$. Diamonds display the profile averaged over four turbulent calculations with $\gamma=10^{-4}$, and time-averaged between 4 and 5 orbits. The solid curve shows the result of Equation~(\ref{vortdirac}), also time-averaged between 4 and 5 orbits, with $D = 4\nu$.}
\end{figure}
The saturation level of the horseshoe drag depends on the advection-diffusion of vortensity inside of the planet's horseshoe region. In laminar disks, vortensity diffusion is controlled by the kinematic viscosity (see \S~\ref{sec:hdd}), while in turbulent disks it should depend on the turbulence strength. In \S~\ref{sec:alpha}, we have related the turbulence amplitude $\gamma$ to an equivalent Reynolds alpha parameter $\langle\alpha_R\rangle$, which quantifies the turbulent transport of angular momentum. There is no reason to expect that the turbulent viscosity associated with $\langle\alpha_R\rangle$, namely the quantity $\nu_R = \langle\alpha_R\rangle c_s H$, should correspond to the vortensity's diffusion coefficient, which we denote by $D$. To evaluate $D$, we altered the unperturbed surface density profile by imposing a slight overdensity ($10\%$) on one ring, located at $r=r_d$. A perturbation of same amplitude, but of opposite sign, is thus applied to the unperturbed vortensity profile. In absence of viscosity and turbulence, the density profile flattens out while the vortensity profile is maintained at its initial value, to within the effects of numerical viscosity. This actually allowed us to check that the numerical viscosity is much smaller than the minimum values of laminar and turbulent viscosities used in our study. For laminar and turbulent runs, we checked that the time evolution of the vortensity profile $V$ (subtracted from the background profile) takes the form
\begin{equation}
V(r,t) = V_d\,\delta r\,\frac{e^{-(r-r_d)^2 / 4 D t}}{\sqrt{4\pi D t}},
\label{vortdirac}
\end{equation}
where $V_d$ is the initial vortensity perturbation at $r=r_d$, and $\delta r$ is the mesh radial resolution. This is illustrated in Figure~\ref{schmidt} for a laminar run with $\nu = 3.4\times 10^{-7}$, and a turbulent run with $\gamma=10^{-4}$. For both calculations, $h_p = 3\%$, $r_d \approx 1.1$, and $2V_d\Sigma_p / \Omega_p = 0.1$. The values of $\nu$ and $\gamma$ were taken such that $\langle \alpha_R \rangle$, given by Equation~(\ref{alpha}), is equal to $\nu / c_s H$. Both calculations thus have the same equivalent Reynolds alpha viscosity: $\langle \alpha_R \rangle \approx 3.8\times 10^{-4}$. The vortensity profile (subtracted from the background profile) is shown at $t=4.5\,T_{\rm orb}$ for the laminar run (stars). The dash-dotted curve is the perturbed profile expected from Equation~(\ref{vortdirac}) with $D=\nu$, at the same time. It shows that, as expected, the vortensity's diffusion coefficient in laminar disks corresponds to the kinematic viscosity \citep[e.g.][]{og2006}. Also displayed is the perturbed vortensity profile time-averaged between 4 and 5 orbits, and further averaged over four different turbulent calculations (diamonds, the turbulent runs differ by the random numbers sorted in the expression for the turbulent potential). The solid curve shows the perturbed profile given by Equation~(\ref{vortdirac}), time-averaged over the above time range, with $D = 4\langle \alpha_R \rangle c_s H = 4\nu$. The good agreement with the numerical profile underscores that, in our turbulent tuns, the vortensity's diffusion coefficient differs from the turbulent viscosity $\nu_R$ associated with the turbulent transport of angular momentum. By considering several other values of $\gamma$, we have estimated the ensemble average of the vortensity's diffusion coefficient as $D \approx 4\nu_R$, with a relative uncertainty of $\approx 30\%$. Differently stated, vortensity diffusion is four times {\it more efficient} than angular momentum transport in our turbulent calculations.

The ratio $\nu_R / D$ is usually referred to as the Schmidt number $S_c$. For the vortensity radial diffusion, we thus find a (radial) Schmidt number of $S_c \approx 0.25$. Similar values, albeit slightly smaller, were obtained with the diffusion of passive scalars. For comparison, studies with 3D MHD calculations report radial and vertical Schmidt numbers for dust diffusion of order unity \citep[see][and references therein]{fn09}. Note that in such simulations, $\nu_R$ is substituted by $\nu_R + \nu_M$, where $\nu_M$ is the viscosity associated with the Maxwell stress tensor. Schmidt numbers are however sensitive to the particles sizes, and they can become much smaller than unity for small particles well-coupled to the gas, thus acting as passive scalars \citep{fn09}.

In laminar disks, the kinematic viscosity $\nu$ is commonly modeled by a dimensionless alpha viscosity $\alpha = \nu / c_s H$. Similarly, we define for turbulent runs an equivalent alpha viscosity $\langle\alpha_D\rangle$ as $\langle\alpha_D\rangle = D / c_s H$. Note that $\langle\alpha_D\rangle = S_c^{-1} \langle\alpha_R\rangle \approx 4\langle\alpha_R\rangle $. Equation~(\ref{alpha}) can thus be recast as
\begin{equation}
	\gamma \approx 8.5\times 10^{-2}\,h_p\,\sqrt{\langle\alpha_D\rangle},
\label{alpha2}
\end{equation}
which inverts as $\langle\alpha_D\rangle \approx 1.4\times 10^{2}(\gamma/h_p)^2$. Interestingly, in MHD calculations, the Maxwell stress parameter $\alpha_M$ is typically three times larger than the Reynolds stress parameter $\alpha_R$ \citep[e.g.][]{fn09}. The total alpha parameter $\alpha$ thus verifies $\alpha \sim 4\alpha_R$. Since the Schmidt number is also commonly of order unity in MHD calculations (but see above), we suggest that our value of $\langle\alpha_D\rangle$, given by Equation~(\ref{alpha2}), can model the total alpha parameter of MHD simulations.

Typical alpha parameters in active zones are in between $10^{-3}$ and $10^{-1}$, depending on the geometry of the magnetic field \citep[see e.g.][]{fromang07a}. Even if the magnetorotational instability only develops in the disk layers, turbulence may be induced in the dead zone \citep{flemingstone03}. These authors estimated that the alpha parameter in dead zones might be about two orders of magnitude smaller than in active layers. Values of $\alpha$ ranging from $10^{-5}$ to $10^{-4}$ might thus be relevant for dead zones \citep[see also][]{oishi07}. From Equation~(\ref{alpha2}), and assuming $h_p=5\%$, we find that values of $\gamma$ in the range $[10^{-4}-10^{-3}]$ should be relevant for active layers, while $\gamma \lesssim 5\times 10^{-5}$ should be more appropriate for dead zones.

\subsection{Running-time average of the turbulent torque}
\label{sec:rta}
\begin{figure}
   \includegraphics[width=\hsize]{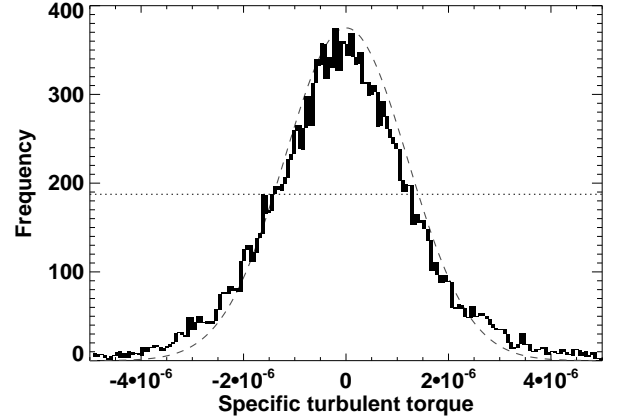}
   \caption{\label{tqrms}Distribution of the specific turbulent torque obtained with $\gamma = 10^{-5}$. A Gaussian distribution centered about zero, with standard deviation $\sigma = 1.2\times 10^{-6}$, is superimposed (dashed curve). The horizontal dotted line corresponds to half the maximum value of the torque distribution.}
\end{figure}
In addition to potentially desaturating the horseshoe drag, the turbulence also exerts a stochastic torque on the planet through the density perturbations it generates. This stochastic torque is referred to as the turbulent torque and is denoted by $\Gamma_{\rm turb}$. We are primarily interested in the cumulative effects of the turbulent torque, which we can evaluate through its running-time average $\overline\Gamma_{\rm turb}$, defined as $\overline\Gamma_{\rm turb}(t) = t^{-1}\times\int_0^t \Gamma_{\rm turb}(u)du$. To investigate the properties of $\overline\Gamma_{\rm turb}$, we performed a series of calculations with a massless planet, and with varying the turbulence coefficient $\gamma$. The specific turbulent torque (turbulent torque by unit planet mass) is directly given by the instantaneous specific torque exerted by the disk on the planet. We display in Figure~\ref{tqrms} the distribution of the specific turbulent torque for $\gamma = 10^{-5}$, after 1000 orbital periods. A Gaussian best fit is overplotted and shows that the turbulent torque frequency follows a Gaussian distribution. Thus, the amplitude of $\overline\Gamma_{\rm turb}$ can be written as \citep[e.g.][]{nelson05}
\begin{equation}
|\overline\Gamma_{\rm turb}| \approx \sigma_{\rm turb} \times \left( \frac{t}{\tau_c} \right)^{-1/2}~~{\rm for}~~t \gg \tau_c,
\label{gammaturb}
\end{equation}
with $\sigma_{\rm turb}$ the mean deviation of the turbulent torque distribution. It takes the form $\sigma_{\rm turb} = C\,\Sigma_p\,q\,\gamma\,r_p^4\,\Omega_p^2$, where $C$ is a dimensionless constant, and where the quantities with a $p$ subscript are to be evaluated at the planet location (LSA04, OIM07). It can be recast as $\sigma_{\rm turb} = \tilde{C}\,2\pi G\Sigma_p M_p r_p$, which corresponds to a fraction $\tilde{C}$ of the natural scale for the torque exerted by the disk on the planet \citep[e.g.][]{jgm06}. The functional dependence of $\sigma_{\rm turb}$ is expected from the expression of the turbulent potential, given by Equation~(\ref{turbpot}). In particular, $\sigma_{\rm turb}$ does not depend on the disk scale height, as in 3D MHD calculations \citep[][]{qmwmhd4}. 

We checked with our series of calculations that $\sigma_{\rm turb}$ scales with $\gamma$. The value of the constant $C$ was inferred from the turbulent torque distributions. We find that:
\begin{equation}
	\sigma_{\rm turb} \approx 2.4\times 10^{2}\,\Sigma_p\,q\,\gamma\,r_p^4\,\Omega_p^2,
	\label{sigmaturb}
\end{equation}
when the $m>6$ modes truncation is taken into account. The above expression should otherwise be multiplied by a factor of $\approx 1.25$. This is about $20$ times larger than the expression given by LSA04. We show in \S~\ref{sec:issues} that such large difference cannot be explained by the absence of $m=1$ modes in the work of LSA04. It is also interesting to compare the expression in Equation~(\ref{sigmaturb}) with the results of 3D MHD calculations. Using Equation~(\ref{alpha2}) together with typical disk parameters in the 3D MHD simulations of \cite{nelson05} -- $\alpha \approx 8\times 10^{-3}$ at $r_p \approx 2.5$, $h_p = 7\%$, and $\Sigma_p r_p \approx 5.2\times 10^{-4}$ \citep{jgm06} -- we find $\sigma_{\rm turb} / q \approx 7\times 10^{-5}$, which is not far from the value found by \cite{nelson05} ($\sim 1.5\times 10^{-4}$ in the same code units, see his figure 14). This general agreement makes us confident that the results obtained with our two-dimensional simulations reproduce the main turbulence properties of realistic 3D MHD calculations. We finally comment that Equations~(\ref{gammaturb}) and~(\ref{sigmaturb}) have been worked out for a massless planet. The turbulent torque actually felt by a low-mass planet may be different, as the disk responses to the planet potential and to the turbulent potential could be coupled.

\subsection{Convergence time}
\label{sec:convtime}
We assume that the torque on the planet is the sum of the differential Lindblad torque $\Delta\Gamma_{\rm LR}$, the horseshoe drag $\Gamma_{\rm HS}$, and the turbulent torque $\Gamma_{\rm turb}$. The results shown in \S~\ref{sec:numres} are notably aimed at testing this assumption. To properly assess the desaturation level of the horseshoe drag with turbulence, we must ensure that $|\overline\Gamma_{\rm turb}|$ is a small fraction of $|\Delta\Gamma_{\rm LR} + \Gamma_{\rm HS}|$. We denote by $f$ this fraction. This constraint provides the convergence timescale $t_{\rm conv}$ of the calculations. From Equation~(\ref{gammaturb}), we have
\begin{equation}
	t_{\rm conv} \approx \tau_c \, f^{-2} \times \left( \frac{\sigma_{\rm turb}}{\Delta\Gamma_{\rm LR} + \Gamma_{\rm HS}} \right)^2.
	\label{tauconv1}
\end{equation}
The differential Lindblad torque is evaluated by a recent estimate derived by \cite{pbck09}. This estimate, which updates the standard formula of \cite{tanaka2002} by including the softening length of the planet potential, reads:
\begin{equation}
	\Delta\Gamma_{\rm LR} = -C_{\rm LR}\,q^2\,\Sigma_p\,r_p^4\,\Omega_p^2\,h_p^{-2},
	\label{dglr}
\end{equation}
with 
\begin{equation}
	C_{\rm LR} \approx (2.5 + 1.7\tau - 0.1\sigma)\times\left(  \frac{0.4H_p}{\varepsilon} \right)^{0.71},
	\label{clr}
\end{equation}
where $\tau = -d\log T_0 / d\log r$ is the power-law index of the unperturbed temperature profile $T_0$, and $H_p = h_p r_p$. Equation~(\ref{dglr}) differs from the 3D estimate of \cite{tanaka2002} by the constant $C_{\rm LR}$, which reads $C_{\rm LR} \approx 2.340 - 0.099\sigma + 0.418\tau$ in the latter work\footnote{We comment that \cite{tanaka2002} provided estimates of the differential Lindblad torque, and of the linear corotation torque for {\it locally} isothermal disks (except for the linear corotation torque in two dimensions). In three dimensions, the total linear torque reads $\Gamma_{\rm 3D} = -C_{\rm 3D}\,q^2\,\Sigma_p\,r_p^4\,\Omega_p^2\,h_p^{-2}$, with $C_{\rm 3D} = 1.364 + 0.541\sigma + 0.433\tau$.}. For isothermal disks ($\tau=0$), both formula give very similar results, as far as conservative values of the softening length are concerned. For $\varepsilon / H_p = 0.6$, they differ by $\sim 25\%$. 

For the sake of simplicity, we assume that the horseshoe drag takes its fully unsaturated value. In isothermal disks, the fully unsaturated horseshoe drag reads \citep{wlpi91}:
\begin{equation}
	\Gamma_{\rm HS} = \frac{3}{4}\left( \frac{3}{2} - \sigma \right)x_s^4\,\Sigma_p\,\Omega_p^2,
	\label{gammahs}
\end{equation}
where $x_s$ is the half-width of the planet's horseshoe region, and where $3/2-\sigma$ is the opposite of the vortensity gradient. By equating the expression in Equation~(\ref{gammahs}) with the two-dimensional linear corotation torque formula of \cite{tanaka2002}, \cite{mak2006} estimated the half-width of the horseshoe region as
\begin{equation}
	x_s \approx 1.1r_p \left(\frac{q}{h}\right)^{1/2}.
	\label{xs_m06a}
\end{equation}
Using a simple two-dimensional model for the horseshoe region, \cite{pp09b} have recently derived an analytic expression for $x_s$ featuring the softening length of the planet's potential (see their equation 39). For small softening ($\varepsilon \ll 0.1H_p$), their expression agrees with the values of $x_s$ measured in numerical simulations, and it differs from the above expression of \cite{mak2006} by a factor of $\sim 1.5$, quite independently of $\varepsilon$ (see their figure 10). For $\varepsilon \gtrsim 0.1H_p$, the analytic expression of \cite{pp09b} is reduced by a fitting factor of $\sim 30\%$, meant to reproduce results of simulations. The resulting estimate is in good agreement with that of \cite{mak2006} for $\varepsilon/H_p$ ranging from $0.2$ to $0.6$. For the run parameters of \S~\ref{sec:numres} ($q/h^3 \approx 0.2$, $\varepsilon = 0.6H_p$), they agree to within $10\%$ \citep[][figures 10 and 11]{pp09b}. They are also in good agreement with our results of simulations, which we checked by a streamline analysis. Using Equation~(\ref{xs_m06a}), Equation~(\ref{gammahs}) can then be recast as
\begin{equation}
	\Gamma_{\rm HS} = -C_{\rm HS}\,q^2\,\Sigma_p\,r_p^4\,\Omega_p^2\,h_p^{-2},
	\label{gammahs2}
\end{equation}
with
\begin{equation}
	C_{\rm HS} \approx -1.1\times \left( \frac{3}{2} - \sigma \right).
	\label{chs}
\end{equation}

Combining Equations~(\ref{alpha2}),~(\ref{sigmaturb}) to~(\ref{clr}),~(\ref{gammahs2}) and~(\ref{chs}), we are left with:
\begin{equation}
	t_{\rm conv} \approx 4\times 10^2\,\tau_c\,\frac{\langle\alpha_D\rangle}{f^{2}\left(C_{\rm LR}+C_{\rm HS}\right)^{2}}\left(  \frac{q}{h_p^3} \right)^{-2},
	\label{tauconv2}
\end{equation}
where we recall that the horseshoe drag $\Gamma_{\rm HS}$ was assumed to take its fully unsaturated value. Equation~(\ref{tauconv2}) may be refined by including the analytic dependence of $\Gamma_{\rm HS}$ with viscosity expected in laminar disks \citep{masset02}. This would come to substituting $C_{\rm HS}$ in Equation~(\ref{chs}) by $C_{\rm HS}\,{\cal F}(z_s)$, where the quantity ${\cal F}(z_s)$ is given by Equation~(\ref{calF1}). When $C_{\rm HS} < 0$, namely for $\sigma < 3/2$, the expression in Equation~(\ref{tauconv2}) corresponds to a {\it maximum} convergence time.

Further insight can be obtained by estimating the range values of $\langle\alpha_D\rangle$ such that the horseshoe drag is unsaturated. As recalled in \S~\ref{sec:hdd}, this condition is roughly fulfilled in laminar disks when the diffusion timescale across the horseshoe region, $\tau_{\rm visc} \approx x_s^2 / \nu$, is smaller than the libration period $\tau_{\rm lib} = 8\pi r_p / (3\Omega_p x_s)$, but larger than the horseshoe U-turn time $\tau_{\rm u-turn} \approx h_p \tau_{\rm lib}$ \citep{bm08a}. Anticipating that this condition essentially holds with turbulence (see \S~\ref{sec:hsregion}), and assuming that $x_s \approx 1.1 r_p (q / h_p)^{1/2}$, we find that the value of $\langle\alpha_D\rangle$ at $r=r_p$ should verify
\begin{equation}
	0.16\,q^{3/2}\,h_p^{-7/2} < \langle\alpha_D\rangle < 0.16\,q^{3/2}\,h_p^{-9/2}.
	\label{alphadesaturation}
\end{equation}
We therefore assume that the horseshoe drag is maintained at its fully unsaturated value for $\langle\alpha_D\rangle \approx 0.16\,q^{3/2}\,h_p^{-4}$. This estimate is checked against numerical simulations in \S~\ref{sec:complam}. Equation~(\ref{tauconv2}) can finally be recast as
\begin{equation}
	t_{\rm conv} \approx 65\,\tau_c\,\frac{q^{-1/2}\,h_p^{2}}{f^{2}\left(C_{\rm LR}+C_{\rm HS}\right)^{2}}.
	\label{tauconv3}
\end{equation}
We emphasize again that the expression given in Equation~(\ref{tauconv3}) assumes that the horseshoe drag is fully unsaturated, and that $C_{\rm HS}$ is independent of $\varepsilon$. For our purposes, minimizing the convergence time implies a compromise between a small aspect ratio, and a planet mass that is not too large to be relevant for type I migration. The disk and planet parameters used in our calculations are described in \S~\ref{sec:numres}.
\begin{figure}
  \centering
	 \includegraphics[width=0.8\hsize]{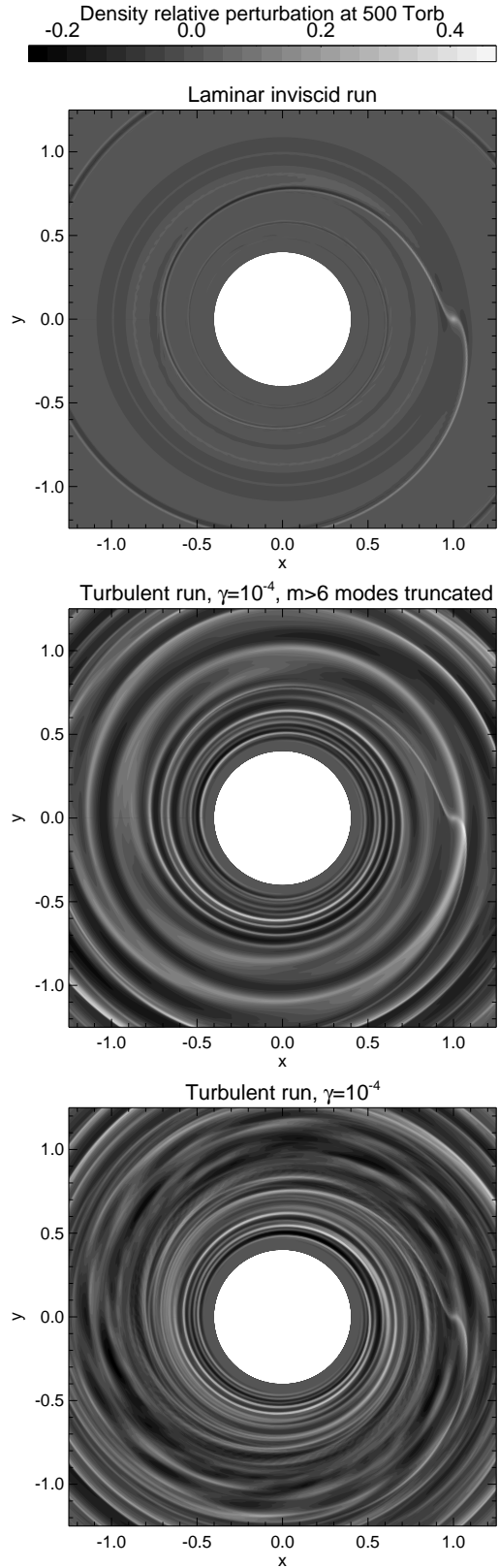}
         \caption{\label{densplots}Relative perturbation of the disk surface density at 500 orbits for $\gamma=0$ (top panel), and $\gamma=10^{-4}$ with and without the $m>6$ modes truncation (middle and bottom panels, $\gamma=10^{-4}$ is equivalent here to $\langle\alpha_D\rangle \approx 1.5\times 10^{-3}$). A $M_p \approx 1.6M_{\oplus}$ planet mass is located at $x=1$, $y = 0$.}
\end{figure}

\section{Results of numerical simulations}
\label{sec:numres}
We present in this section the results of simulations with planet, and with the turbulence model detailed in \S~\ref{sec:turb}. These simulations follow the numerical set-up described in \S~\ref{sec:setup}. Recall in particular that the power-law index of the initial surface density profile is $\sigma = 0.5$. The disk and planet parameters were chosen to minimize the maximum convergence time $\tau_{\rm conv} $ of the turbulent runs, given by Equation~(\ref{tauconv3}). We took $h_p = 3\%$ and $q = 5\times 10^{-6}$, which corresponds to a $M_p \approx 1.6M_{\oplus}$ planet mass if the central object has a solar mass. Using $\tau_c \approx 0.5$ orbits (see \S~\ref{sec:act}), and assuming $f=0.1$, we have $\tau_{\rm conv} \sim 2000$ orbits. Our simulations were run for about 4000 orbits, so we expect that $|\overline\Gamma_{\rm turb}|$ should not exceed $\sim 0.1 |\Delta\Gamma_{\rm LR} + \Gamma_{\rm HS}|$ at the end of the simulations. For comparison, laminar runs without turbulence, but with a constant kinematic viscosity $\nu$ are performed. Runs with turbulence required about twice as much computing time as runs without turbulence. We also point out that the half-width of the planet's horseshoe region is $x_s \approx 0.015r_p$. It is resolved by about 6 cells along the radial direction. This resolution is similar to that of other recent numerical studies of the horseshoe drag. The influence of the grid resolution on our results is assessed in \S~\ref{sec:issues}.

\subsection{Torque vs. turbulence amplitude}
\label{sec:desaturation}
\begin{figure*}
   \includegraphics[width=0.5\hsize]{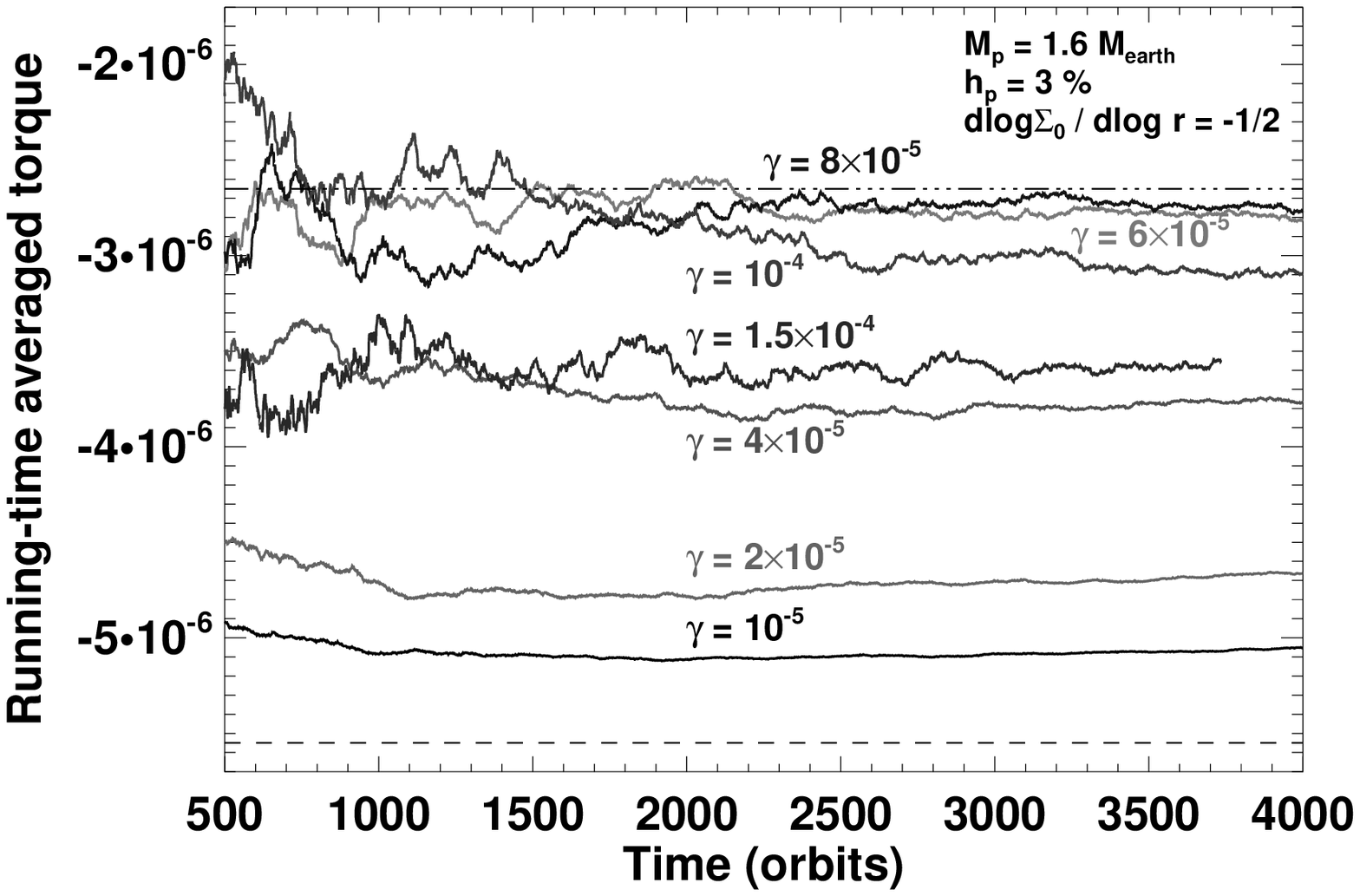}
   \includegraphics[width=0.5\hsize]{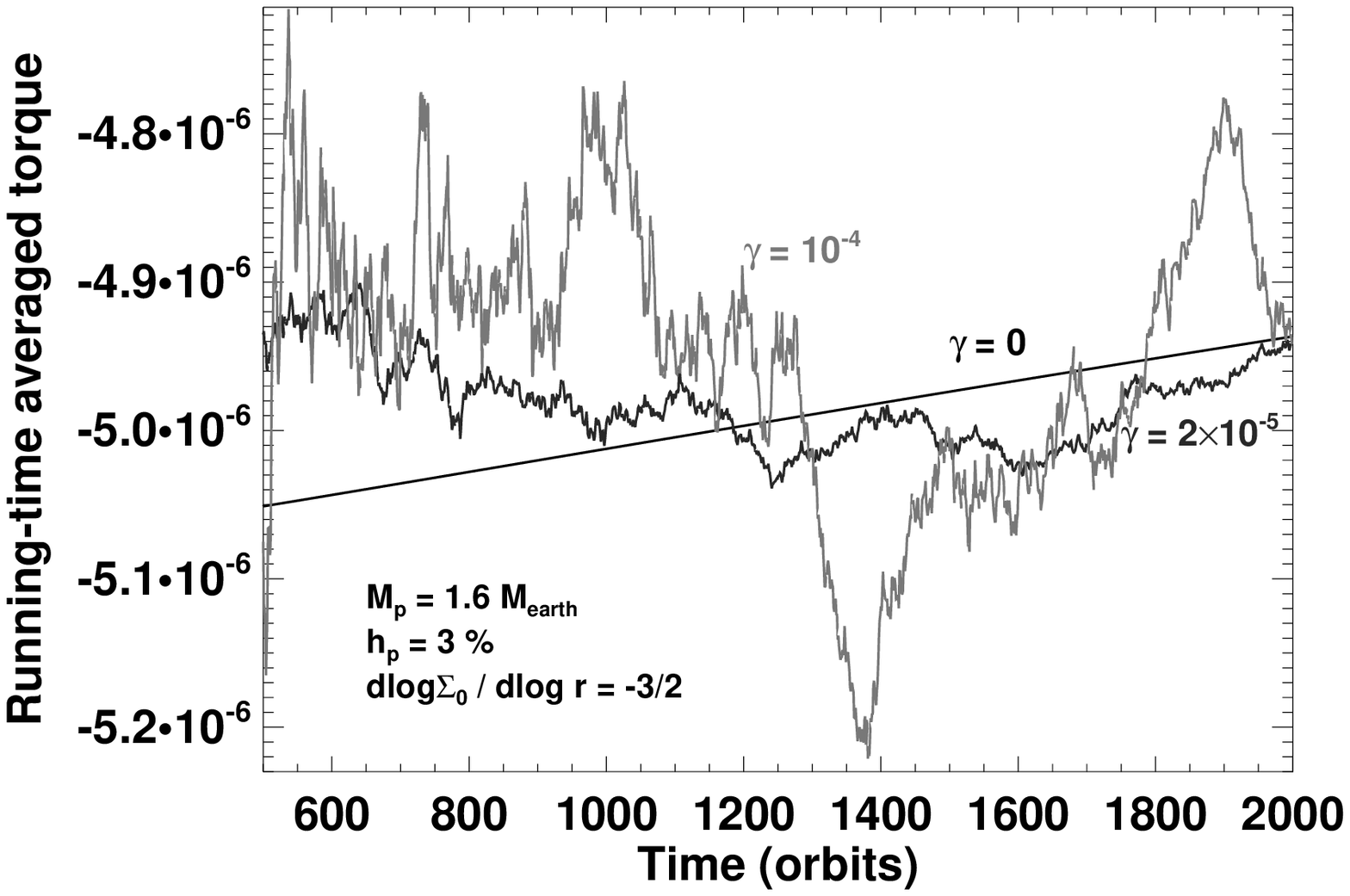}
  \caption{\label{rta}Time evolution of the running-time averaged specific torque exerted on a $M_p \approx 1.6M_{\oplus}$ planet mass. Results are shown for several values of $\gamma$, for $\sigma = 0.5$ (left panel) and $\sigma = 1.5$ (right panel). The dashed and dash-dotted lines in the left panel depict the values of the differential Lindblad torque and of the fully unsaturated torque, respectively, both measured in a run without turbulence.}
\end{figure*}
We performed a series of 9 runs with varying the turbulence amplitude from $\gamma=10^{-5}$ to $\gamma=1.5\times 10^{-4}$. Using Equation~(\ref{alpha2}), the equivalent alpha viscosity $\langle\alpha_D\rangle$ related to the vortensity's turbulent diffusion ranges from $1.5\times 10^{-5}$ to $3.5\times 10^{-3}$. A laminar inviscid run ($\gamma=0$, $\nu=0$) was performed for comparison. The top panel of Figure~\ref{densplots} displays the relative perturbation of the disk's surface density (with respect to the initial profile) obtained at 500 orbits for the laminar inviscid run. A shallow gap is progressively opened up, which is consistent with the non-linear wave dissipation model of \cite{rafikov02} in inviscid disks. This model predicts that the minimum planet to primary mass ratio $q_{\rm min}$ for gap opening is
\begin{equation}
q_{\rm min} = \frac{2h_p^3}{3} \times {\rm Min} \left\{ 5.2Q_p^{-5/7}, 3.8\left(Q_p/h_p\right)^{-5/13} \right\}.
\label{qminrafikov}
\end{equation}
The Toomre parameter at the planet location is $Q_p \sim 20$, and Equation~(\ref{qminrafikov}) gives $q_{\rm min} \approx 5.7\times 10^{-6}$. This value is in agreement with our planet to primary mass ratio. Note however that this model does not include the planet's softening length, and the possible interaction between the planet wake and the flow inside of the horseshoe region \citep{mak2006}. We mention that a shallow gap also formed for the lowest values of the turbulence amplitude. The gap impact on the torque variation remains small, as will be shown below. The middle panel of Figure~\ref{densplots} displays the perturbed density for the turbulent run with $\gamma = 10^{-4}$, which corresponds to $\langle\alpha_D\rangle \approx1.5\times 10^{-3}$. For comparison, the perturbed density obtained with $\gamma = 10^{-4}$, but without the $m>6$ modes truncation, is shown in the bottom panel of Figure~\ref{densplots}. In both cases, the planet wake and the density perturbations due to turbulence have similar amplitudes, and no gap is visible. For these runs, the averaged Mach number associated with the velocity's turbulent perturbations is $\approx 0.3$. We comment that a similar wave pattern is obtained in absence of any embedded planet. Note that the spiral waves generated by our turbulence model are more tightly wound close to the disk's inner edge, since the modes radial extent scales with radius. 

The time history of the running-time averaged torques (hereafter, r.t.a. torques) obtained with our series of runs are depicted in the left panel of Figure~\ref{rta}. The torques seem to be all converged with time by the end of the simulations. The dashed and dash-dotted lines show respectively the values of the fully saturated and unsaturated torques, both measured in the run without turbulence. We comment that the fully saturated torque experiences a slow, stationary increase due to the continuous building up of a gap around the planet. For our set of planet and disk parameters, the fully saturated torque only increases by $\sim 4\%$ between $500$ and $4000$ orbits. It can thus be confounded with the differential Lindblad torque. As $\gamma$ increases, the stationary value of the r.t.a. torque increases from the differential Lindblad torque up to the fully unsaturated torque (here for $\gamma \approx 8\times 10^{-5}$), before it decreases with $\gamma$. This behavior reminds us of the torque dependence with viscosity in laminar disks \citep{masset02}. This result is a strong indication that turbulence can unsaturate the horseshoe drag, depending on the turbulence strength.

To further investigate this result, we performed another series of calculations with an initial surface density profile decreasing as $r^{-3/2}$. In laminar isothermal disks, the horseshoe drag cancels out at all time, and the total torque is equal to the differential Lindblad torque. The results of these simulations are displayed in the right panel of Figure~\ref{rta}, for $\gamma=0$, $\gamma=2\times 10^{-5}$ and $\gamma=10^{-4}$. Running-time averaged torques are depicted, even for the run without turbulence. For this run, recall that the slow increase of the torque is due to the gap clearance. The convergence time is clearly much smaller than in the previous series of runs with $\sigma=0.5$. Using Equations~(\ref{alpha2}) and~(\ref{tauconv2}), and assuming $f=0.1$, we have $\tau_{\rm conv} \sim 300\,T_{\rm orb}\,(\gamma / 10^{-4})^2$. For the values of $\gamma$ considered here, $|\overline\Gamma_{\rm turb}|$ should be already smaller than $\sim 0.1 |\Delta\Gamma_{\rm LR}|$ from $\sim 500$ orbits, which is clearly seen in the right panel of Figure~\ref{rta}. The torques with turbulence rapidly converge toward the torque without turbulence. This result could be somewhat surprising since for the largest values of $\gamma$ of our study, the density perturbations due to turbulence are typically as strong as those triggered by the planet wake (see the middle and bottom panels of Figure~\ref{densplots}). It indicates that, for our range values of $\gamma$, and for the duration of our runs, the differential Lindblad torque is {\it not significantly altered} by turbulence, and that the torque on the planet can thus be decomposed into a stationary component (the differential Lindblad torque) and a fast-varying component (the turbulent stochastic torque), with no significant coupling between both. In addition, it justifies that the torque variation with $\gamma$ obtained with $\sigma=0.5$ does arise from the horseshoe drag desaturation through turbulence. More insight into the long-term evolution of runs at high-turbulence is provided in \S~\ref{sec:complam}.

\begin{figure*}
   \includegraphics[width=0.5\hsize]{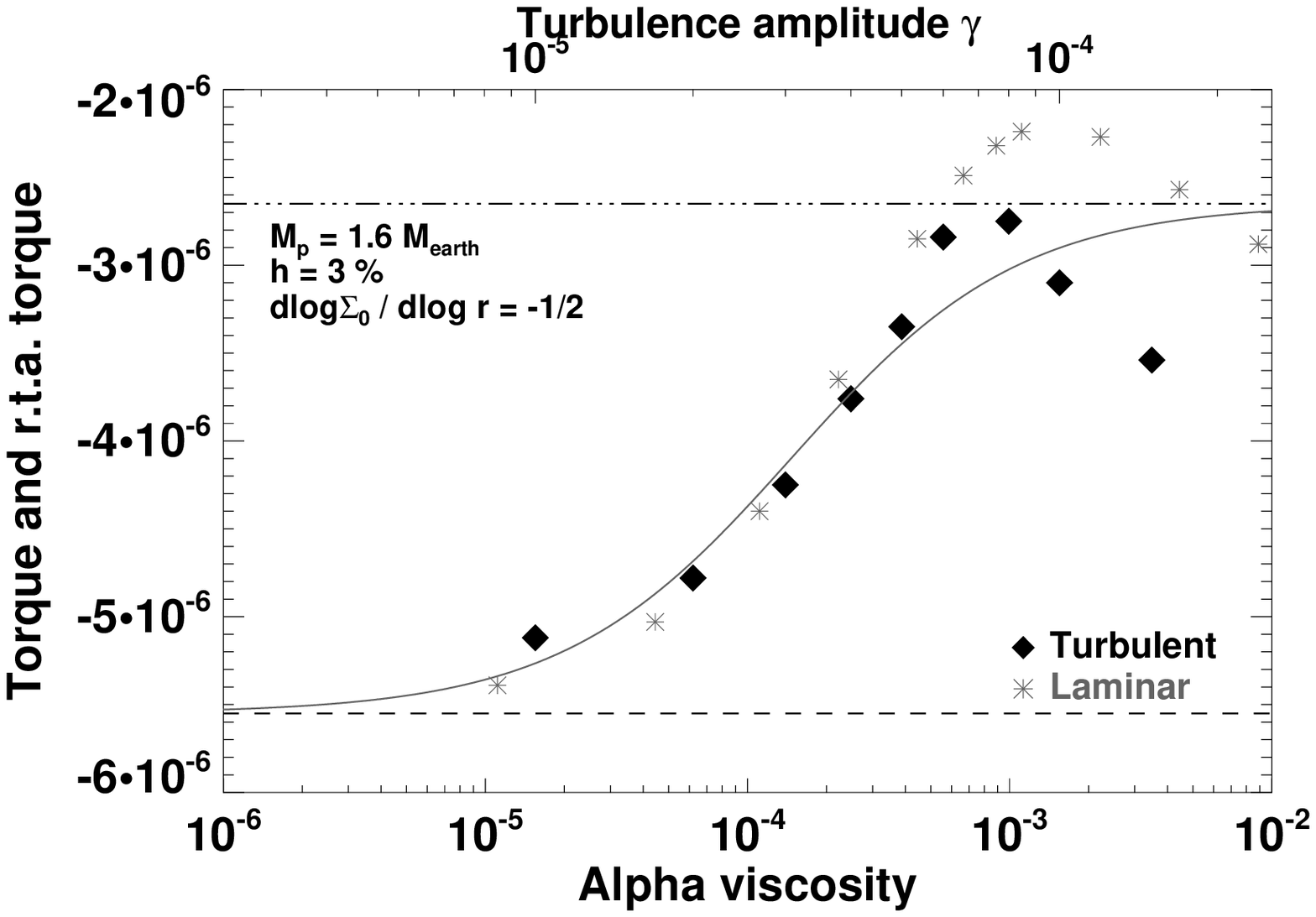}
   \includegraphics[width=0.5\hsize]{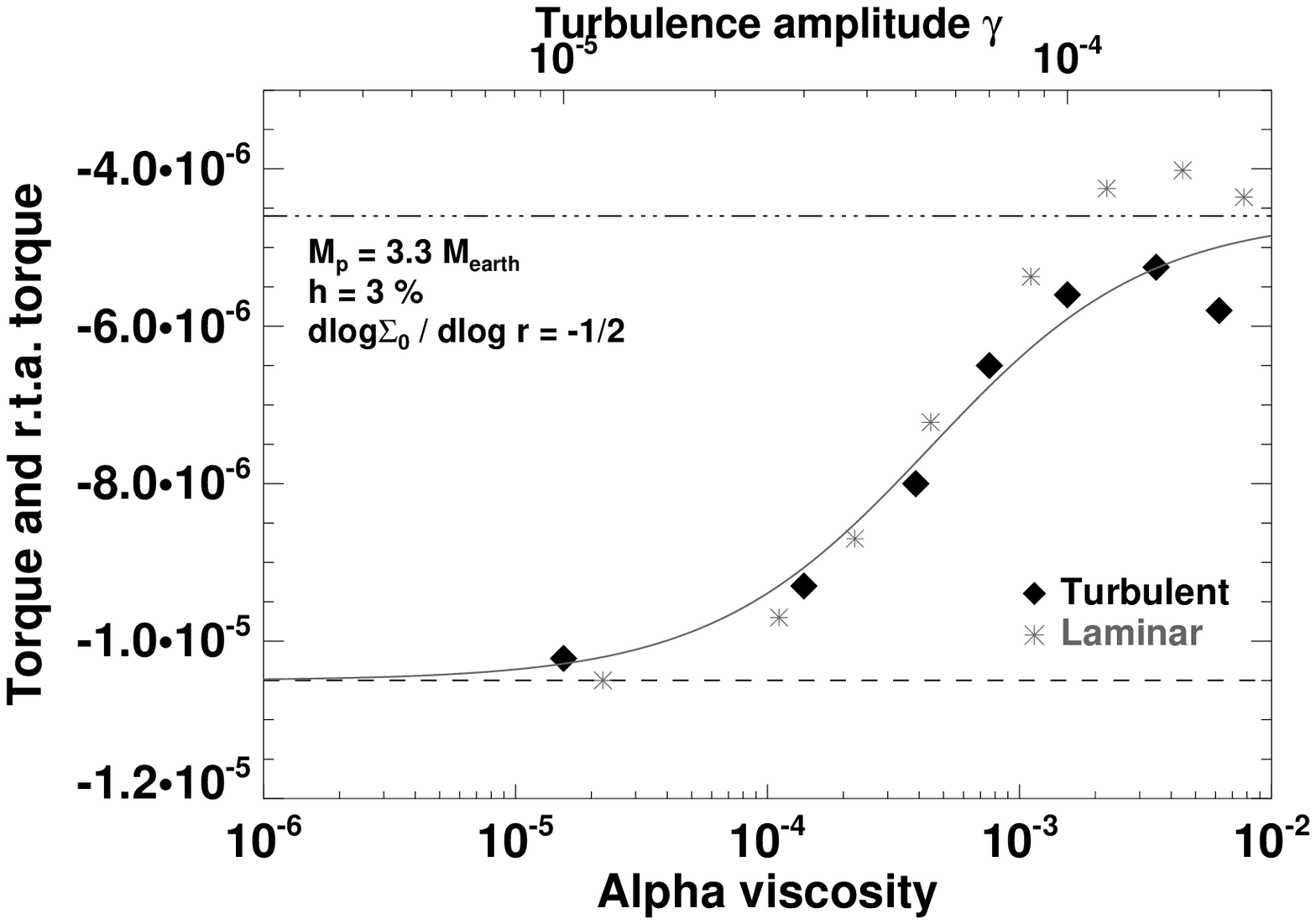}
   \caption{\label{finalrta}Comparison of the turbulent and laminar runs for two planet to primary mass ratios: $q=5\times 10^{-6}$ (left panel, turbulent torques are averaged over 4000 orbits) and $q=10^{-5}$ (right panel, turbulent torques are averaged over 2000 orbits). The alpha viscosity in bottom $x-$axis corresponds to $\alpha$ for laminar runs, and to $\langle\alpha_D\rangle$ for turbulent tuns (see text). The top $x-$axis shows the turbulence amplitude $\gamma$, related to $\langle\alpha_D\rangle$ through Equation~(\ref{alpha2}). The $y-$axis shows the stationary values of the specific torque for laminar runs, and of the specific torque running-time average for turbulent runs. The dashed and dash-dotted lines depict the values of the differential Lindblad torque and of the fully unsaturated torque for an inviscid laminar run. The solid curve displays the dependence of the steady torque with $\alpha$ expected in laminar disks, given by Equation~(\ref{calF1}).}
\end{figure*}

\subsection{Comparison to laminar runs}
\label{sec:complam}
We have shown in \S~\ref{sec:desaturation} that turbulence can unsaturate the horseshoe drag. We now study how the torque dependence with the turbulence amplitude compares to that of the torque with viscosity in laminar disks. For this purpose, we performed an additional series of laminar runs with $\sigma=0.5$ and a uniform kinematic viscosity $\nu$, which can be related to a dimensionless alpha viscosity $\alpha = \nu / c_s H$. Recall that the vortensity's diffusion coefficient $D$ in turbulent runs is similarly modeled by an equivalent alpha viscosity $\langle\alpha_D\rangle = D / c_s H$. We shall assume hereafter that $\alpha$ and $\langle\alpha_D\rangle$ take their value at the planet location. For laminar runs, $\alpha$ varies from $1.1\times 10^{-5}$ to $8.9\times 10^{-3}$. A steady state was attained between 300 to 500 orbits. Stationary torque values obtained with the laminar runs are plotted against $\alpha$ in the left panel of Figure~\ref{finalrta}. Stationary running-time averaged torque values obtained with the turbulent runs are also depicted against $\gamma$ (top axis) and $\langle\alpha_D\rangle$ (bottom axis, same scale as for $\alpha$). 

The results of turbulent and laminar runs are globally in good agreement, especially for intermediate viscosity values. Slight differences at low viscosities are presumably due to the fact that torques were evaluated at different times (from $300$ to $500$ orbits for laminar runs, between $3500$ and $4000$ orbits for turbulent runs). For the smallest viscosities, the r.t.a. torques are thus biased toward more positive values, triggered by the progressive clearance of a gap around the planet. This bias decreases with increasing viscosity. Note that the determination uncertainty in the relationship between $\gamma$ and $\langle\alpha_D\rangle$, given by Equation~(\ref{alpha2}), is another possible source of discrepancies. The overall good agreement between laminar and turbulent calculations {\it at low-turbulence level} implies that the total torque felt by a low-mass planet can be decomposed into a laminar torque and a stochastic torque due to turbulence.

As in Figure~\ref{rta}, the dashed and dash-dotted lines show the values of the differential Lindblad torque $\Delta\Gamma_{\rm LR}$, and of the fully unsaturated torque $\Gamma_{\rm FU}$, both evaluated for $\gamma=0$ and $\nu=0$. The solid curve depicts the analytic dependence of the steady laminar torque with viscosity, given by \cite{masset01}. More precisely, we display the function $\Gamma(z_s)$ defined by
\begin{equation}
\Gamma(z_s) = \Delta\Gamma_{\rm LR} + (\Gamma_{\rm FU} - \Delta\Gamma_{\rm LR}){\cal F}(z_s), 
\label{calF1}
\end{equation}
where
\begin{equation}
{\cal F}(z_s) = 4\left( z_s^{-3} - \frac{g(z_s)}{g^{'}(z_s)}\,z_s^{-4}  \right),
\label{calF2}
\end{equation}
with $z_s = x_s(2\pi\alpha h_p^2 r_p^3)^{-1/3}$. In Equation~(\ref{calF2}), the function $g$ is defined as $g(z) = {\rm Bi}(z) - \sqrt{3}{\rm Ai}(z)$, where ${\rm Ai}$ and ${\rm Bi}$ denote the Airy functions. The half-width of the horseshoe region was determined numerically with the laminar run with $\alpha = 1.1\times 10^{-4}$ (dichotomic search of the separatrices, $x_s$ calculated as the geometric average of the horseshoe half-widths at $\pm 1$ rad, as in \cite{cm09}). We found $x_s \approx 1.429\times 10^{-2}$, which agrees to less than $1\%$ with the estimate $x_s \approx 1.1 r_p (q/h_p)^{1/2}$ of \cite{mak2006}. The analytic expression of Equation~(\ref{calF1}) correctly reproduces the results of both the laminar and turbulent runs, for alpha viscosities smaller than $\sim \alpha_{\rm max}$, where stationary torques take their maximum value.

The differences between laminar and turbulent runs are more evident at high viscosities. Interestingly, laminar runs with $\alpha$ in the range $[5\times 10^{-4} - 5\times 10^{-3}]$ have steady torques that exceed the fully unsaturated torque obtained without viscosity. For any viscosity in this range, the torque excess appears after a time comparable to the horseshoe U-turn time (which indicates that the torque excess is likely an excess of horseshoe drag), and it reaches a maximum value in about a libration time. The maximum laminar torque is obtained for $\alpha_{\rm max} \approx 2\times 10^{-3}$, which is in very good agreement with the estimate $\langle\alpha_D\rangle \approx 0.16\,q^{3/2}\,h_p^{-4}$ used to evaluate the convergence time at Equation~(\ref{tauconv3}). The relative difference between the maximum laminar torque, and the fully unsaturated inviscid torque is $\approx 22\%$. Additional series of calculations\footnote{For these additional runs, we increased the resolution to keep the same ratio $x_s / \delta r$, where $\delta r$ is the mesh size along the radial direction.} revealed that this relative difference decreases with increasing aspect ratio: it amounts approximately to $10\%$ for $h=5\%$, and to $7\%$ for $h=7\%$. A look at the horseshoe drag formulation of \cite{cm09} (their equation 22) suggests that the torque excess can arise from a modification of the horseshoe streamlines (e.g. a slight shift of the separatrices), or of the profiles of vortensity inverse $\Sigma/\omega$ at the upstream parts of the horseshoe region \citep[see e.g.][their figure 2]{mc09}. We performed a thorough streamline analysis that showed no significant modification of the horseshoe streamlines with varying viscosity. The profiles of $\Sigma/\omega$ at the upstream parts of the horseshoe region are modified, however, in a way that we find to be consistent with the presence of a torque excess. The modification of these profiles with viscosity, and therefore the mechanism that triggers the torque excess, deserves a detailed study, which goes beyond the scope of this paper. 

\begin{figure*}
    \includegraphics[width=0.5\hsize]{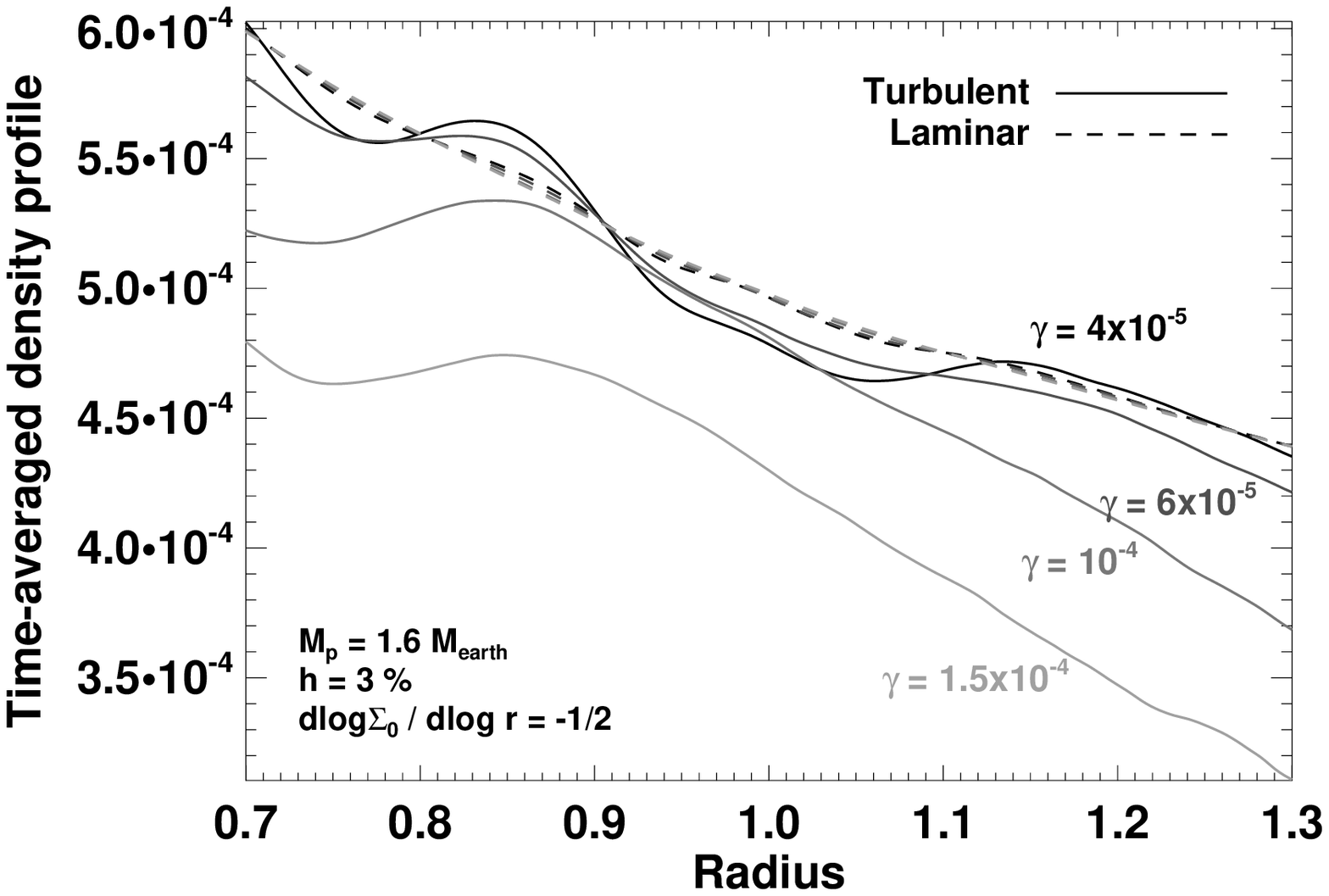}
    \includegraphics[width=0.5\hsize]{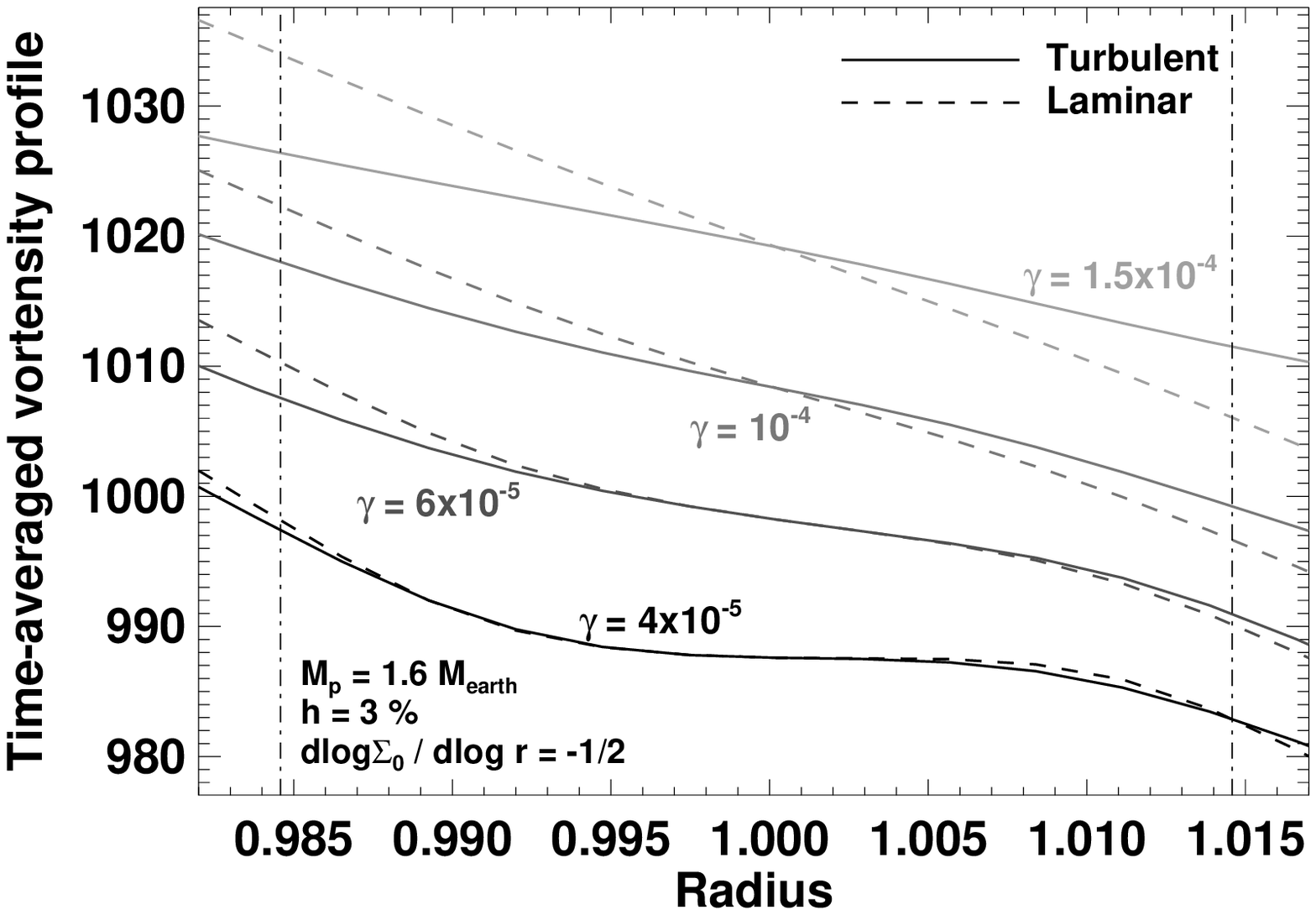}
    \caption{\label{longterm}Surface density profiles (left panel) and vortensity profiles (right panel) for some of the turbulent and laminar runs displayed in the left panel of Figure~\ref{finalrta}. In both panels, profiles with turbulence are time-averaged over $4000$ orbits (solid curves). Dashed curves show the profiles obtained at $300$ orbits with laminar runs. For a fixed value of $\gamma$, laminar and turbulent runs have the same vortensity's diffusion coefficient. In the right panel, all vortensity profiles have been slightly offset to facilitate the comparison of their slopes. The vertical dash-dotted lines show the location of the separatrices of the horseshoe region.}
\end{figure*}
Turbulent runs do not feature such a torque excess. Torques with turbulence start decreasing at slightly smaller viscosities compared to laminar runs, and they seem to decrease faster. To provide more insight into these differences, we show in the left panel of Figure~\ref{longterm} the surface density profiles, time-averaged over $4000$ orbits, for some of the turbulent runs (solid curves). Density profiles obtained at $300$ orbits for laminar runs with similar values of the vortensity's diffusion coefficient are overplotted as dashed curves. At low-viscosity, laminar profiles are not strictly stationary, since a shallow gap ultimately forms around the planet, as can be seen in the turbulent run with $\gamma = 4\times 10^{-5}$. As already shown, the gap's impact on our results is weak, and we neglect it in the following. As $\gamma$ increases, profiles with turbulence tend to steepen around the planet, whereas laminar profiles almost coincide (and will stay so on the long term for the largest viscosities). Up to $\gamma = 10^{-4}$, laminar and turbulent profiles take similar values around the planet, which indicates that the time-averaged Lindblad torque with turbulence, $\Delta\Gamma_{\rm LR}^{\rm T}$, should approximately equal that of the laminar run, $\Delta\Gamma_{\rm LR}^{\rm L}$. For $\gamma = 1.5\times 10^{-4}$, the averaged density profile is reduced by $\approx 15\%$ with turbulence, which should decrease the Lindblad torque accordingly, as the slope of the density profile has little impact on the Lindblad torque. To now estimate the impact of the density change with turbulence on the horseshoe drag, we show in the the right panel of Figure~\ref{longterm} the vortensity profiles for previous runs (time-averaged profiles for turbulent runs, and stationary profiles for laminar runs). They are depicted in a narrow range around the horseshoe region, the location of its separatrices are indicated as dash-dotted lines. Also, all profiles have been slightly offset to facilitate the comparison of their slopes. For $\gamma = 4\times 10^{-5}$, turbulent and laminar vortensity profiles hardly differ, which justifies that the total torques are in very good agreement. Nonetheless, for higher values of $\gamma$, time-averaged vortensity gradients are smaller with turbulence, by about $15\%$, $25\%$, and $45\%$ for $\gamma = 6\times 10^{-5}$, $\gamma = 10^{-4}$, and $\gamma = 1.5\times 10^{-4}$, respectively. For $\gamma = 6\times 10^{-5}$, because the averaged density at the planet location is similar in turbulent and laminar runs, the averaged horseshoe drag with turbulence, $\Gamma_{\rm HS}^{\rm T}$, should be $\approx 15\%$ smaller than the stationary horseshoe drag of the laminar run, $\Gamma_{\rm HS}^{\rm L}$. Using the left panel of Figure~\ref{finalrta}, it turns out that, if the torque difference between laminar and turbulent runs was only due to the density change with turbulence, the averaged total torque with turbulence would be $\approx \Delta\Gamma_{\rm LR}^{\rm L} + 0.85\Gamma_{\rm HS}^{\rm L} = -3.06\times 10^{-6}$, which is about $7\%$ larger than its actual value. Similarly, for $\gamma = 10^{-4}$, the torque with turbulence would be $\approx \Delta\Gamma_{\rm LR}^{\rm L} + 0.75\Gamma_{\rm HS}^{\rm L} = -3.03\times 10^{-6}$, which differs from its actual value by about $3\%$. For $\gamma = 1.5\times 10^{-4}$, the horseshoe drag is altered by the change of the vortensity gradient and of the density at the planet location. Again, if the difference between laminar and turbulent torques arose from the density time-evolution with turbulence, the total torque with turbulence would be $\approx 0.85\times (\Delta\Gamma_{\rm LR}^{\rm L} + 0.55\Gamma_{\rm HS}^{\rm L} ) = -3.27\times 10^{-6}$, which is only $\approx 8\%$ smaller than its actual value. Note that, for $\gamma = 1.5\times 10^{-4}$, the reductions of the (positive) horseshoe drag and of the (negative) Lindblad torque almost compensate, which conspires to make the running-time averaged torque a remarkably stationary quantity. From the above comparison, we conclude that the torque differences at high-turbulence between turbulent and laminar runs can be accounted for by the time-evolution in turbulent runs of the surface density profile around the planet location.

\begin{figure*}
    \includegraphics[width=0.33\hsize]{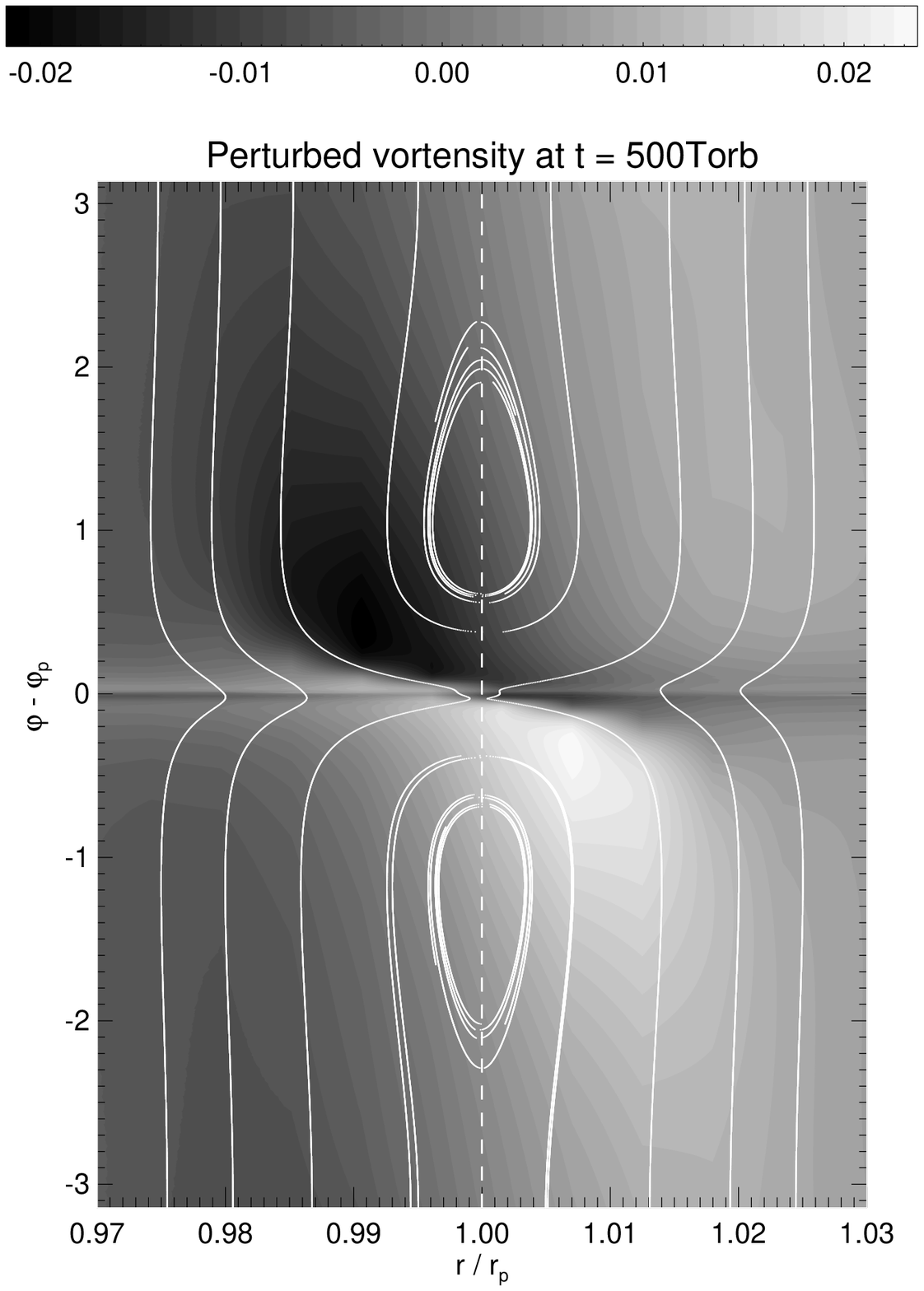}
    \includegraphics[width=0.33\hsize]{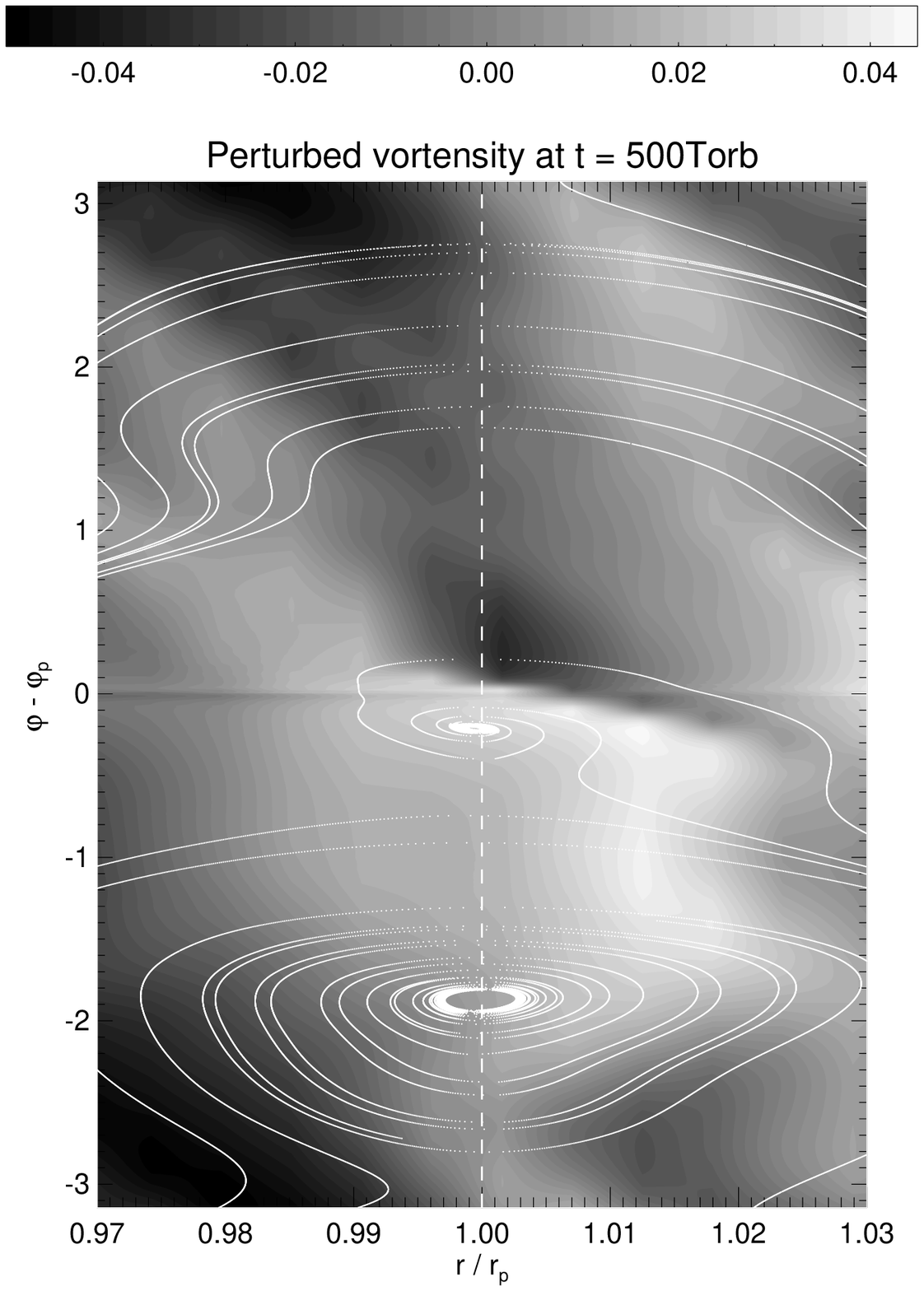}
    \includegraphics[width=0.33\hsize]{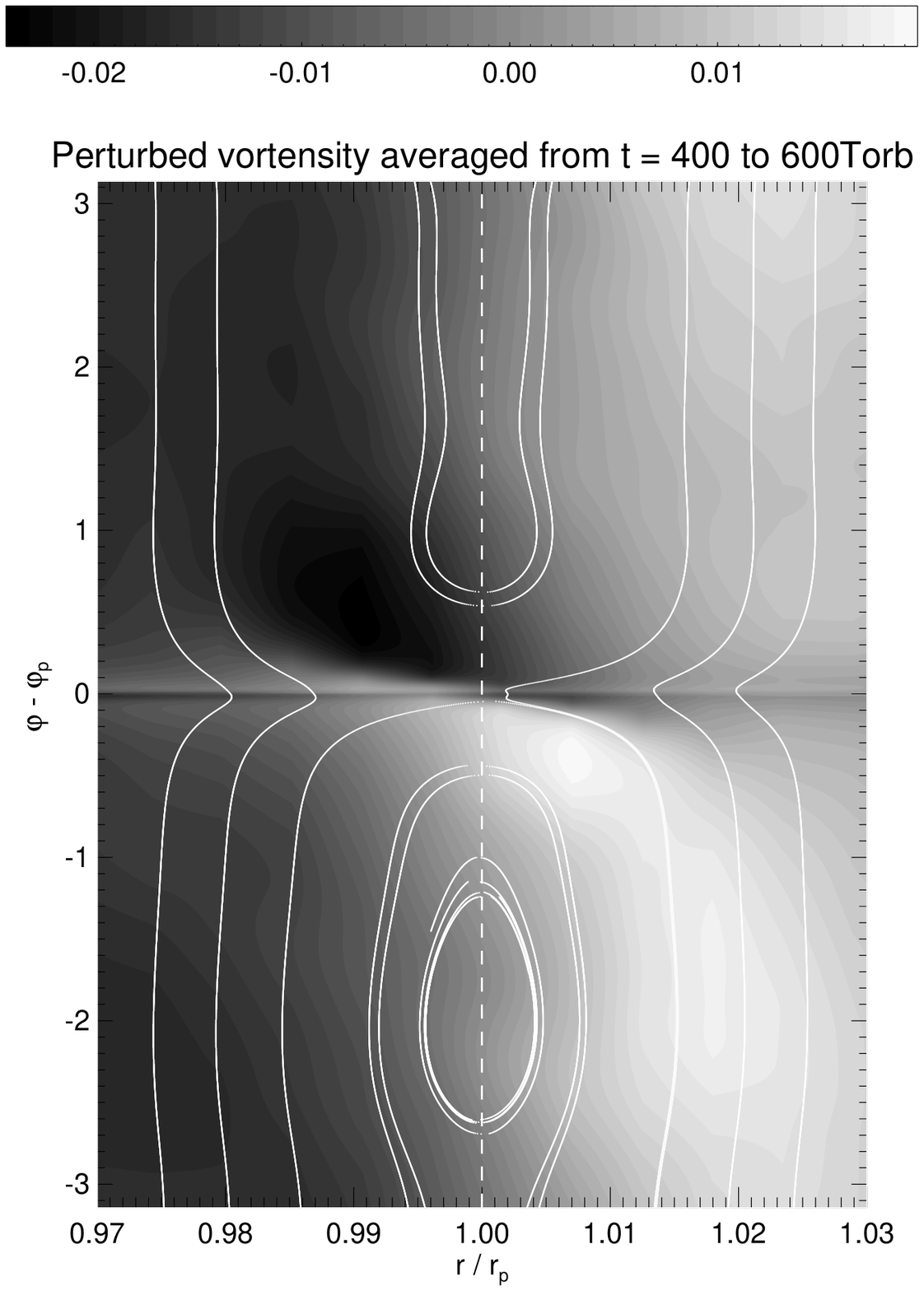}
    \caption{\label{vortplots}Relative perturbation of the disk vortensity (with respect to its initial profile) for a laminar and a turbulent runs. Both calculations have $q=10^{-5}$, $h_p = 5\%$, and $\sigma=0$. Contours are shown in a narrow ring around the planet location. The left panel displays the instantaneous perturbed vortensity at 500 orbits for the laminar run, for which $\alpha = 2\times 10^{-4}$ ($\nu = 5\times 10^{-7}$). Streamlines are overplotted as solid curves. The middle panel shows the instantaneous perturbed vortensity at 500 orbits for the turbulent run, for which $\langle\alpha_D\rangle \sim 2\times 10^{-4}$ ($\gamma = 6\times 10^{-5}$). Instantaneous streamlines are also depicted as solid curves. The right panel plots the perturbed vortensity of the turbulent run, time-averaged between 400 and 600 orbits. Streamlines averaged over the same time period are overplotted. They are directly comparable to those of the left panel. In all panels, the vertical dashed line depicts the planet's orbital radius.}
\end{figure*}
The time-evolution of the surface density profile with turbulence signifies that the initial smooth $r^{-1/2}$ profile is not the equilibrium density profile for our turbulence model. The equilibrium profile departs significantly from a simple power-law profile, as suggested by the flat density transition around $r=0.8$ at high-turbulence. However, since we observe a tendency for the time-averaged density profile to slowly steepen around the planet location, one may wonder whether the density profile could tend toward a $r^{-3/2}$ profile, which would yield a uniform vortensity equilibrium profile. To verify this trend, we performed several calculations with an initial $r^{-3/2}$ surface density profile, at high-turbulence ($\gamma > 3\times 10^{-4}$, corresponding to $\langle\alpha_D\rangle > 1.4\times 10^{-2}$). After about 3000 orbits, we find this time that the time-averaged density profile significantly flattens out. It is possible that the equilibrium density profile is between the $r^{-1/2}$ and $r^{-3/2}$ profiles, and we have not reached yet this equilibrium. In any case, it is likely that turbulence will tend to structure the averaged density profile, due to local variations in the turbulent stress. A numerical artifact due to our set of boundary conditions also cannot be ruled out. Recall that in all of our calculations, wave-damping zones are used along the inner and outer edges of the disk, to avoid reflections. This set of boundary conditions leads to slowly depleting the disk mass outside of the damping zones, in a timescale that depends on the turbulence strength. It explains why, with increasing $\gamma$, the time-averaged density profile with turbulence takes smaller values around the planet. For the highest values of $\gamma$ that we considered, we found that forcing all fields in the damping zones toward their initial value, as in \cite{valborro06}, or toward their instantaneous axisymmetric value, has little impact on the density profile around the planet.

Similar results were obtained with a planet mass twice as large ($q = 10^{-5}$), as shown in the right panel of Figure~\ref{finalrta}. Note that these runs were performed over $2000$ orbits, and not over $4000$ orbits as in previous series with $q = 5\times 10^{-6}$. At a given value of $\gamma$, differences between laminar and turbulent density profiles are thus smaller with $q = 10^{-5}$ than with $q = 5\times 10^{-6}$, and so are the total torques. In addition, two difficulties arise with $q = 10^{-5}$. First, the gap clearance occurs much faster. For instance, the fully saturated torque of the laminar run with the smallest viscosity ($\alpha \approx 2\times 10^{-5}$) increases by $\approx 10\%$ between 400 and 1000 orbits. This increasing rate is about 10 times larger than for an equivalent laminar run with $q = 5\times 10^{-6}$. For the smallest viscosities, for laminar and turbulent runs, torques were thus evaluated before the systematic increase due to the gap opening. Furthermore, the torque of the laminar inviscid run experiences large-amplitude, fast oscillations after $\sim 600$ orbits. These oscillations are triggered by the formation of vortices flowing along the edges of the planet gap \citep{li2005}. The amplitude of the oscillations is several times larger than the differential Lindblad torque, and their period is comparable to the planet's orbital period. We do not get any vortices and torque oscillations in the laminar run with the smallest viscosity ($\alpha \approx 2\times 10^{-5}$).

\subsection{Structure of the horseshoe region with turbulence}
\label{sec:hsregion}
We have shown in \S~\ref{sec:complam} that, when it is time-averaged over a sufficiently long time period, the torque evaluated in a turbulent model is in very good agreement with the torque obtained with a similar laminar model, providing both models have same vortensity's diffusion coefficients ($\alpha \approx \langle\alpha_D\rangle$ with previous notations). Under these precautions, the mean saturation levels of the horseshoe drag in turbulent and laminar calculations are therefore very close. This agreement indicates that, in time-average, the properties of the horseshoe region (width, vortensity advection-diffusion) should be similar in both cases. To investigate these properties, we performed a laminar and a turbulent calculations with $q=10^{-5}$ and $\sigma=0$. The disk aspect ratio at the planet location was increased to $h_p=5\%$, so that the strong vortensity perturbations induced by the planet wake are located outside of the horseshoe region. The laminar run has a kinematic viscosity $\nu=5\times 10^{-7}$ (in code units), which is equivalent to $\alpha = 2\times 10^{-4}$ at the planet location. Similarly, the turbulent simulation was performed with $\gamma = 6\times 10^{-5}$, which corresponds to $\langle\alpha_D\rangle \approx 2\times 10^{-4}$ from Equation~(\ref{alpha2}). This viscosity value is low enough so that the surface density profiles of the turbulent and laminar runs are not significantly altered over the duration of these runs (600 orbits).

We display in the left panel of Figure~\ref{vortplots} the relative perturbation of the vortensity field (with respect to its initial profile) for the laminar calculation, at $500$ orbits. Overplotted as solid curves are streamlines in the planet's frame. The two particular streamlines passing very close to the planet location ($r=r_p$, $\varphi=\varphi_p$) are the horseshoe separatrices. The vertical dashed line shows the location of the planet's corotation radius $r_c$, where $\Omega(r_c) = \Omega_p$. It also corresponds to the planet's orbital radius $r_p$, as the unperturbed pressure profile is uniform for these runs. Since the unperturbed vortensity profile decreases with radius, vortensity advection-diffusion yields negative vortensity perturbations along inward downstream streamlines ($\varphi > \varphi_p$, $r < r_c$), and positive vortensity perturbations along outward downstream streamlines ($\varphi < \varphi_p$, $r > r_c$). These vortensity perturbations decrease in a timescale comparable to the viscous diffusion time across the horseshoe region. As can be seen in the left panel of Figure~\ref{vortplots}, vortensity perturbations cancel out over a time $\sim \tau_{\rm lib}/4$ proceeding horseshoe U-turns.

The middle panel of Figure~\ref{vortplots} shows the instantaneous vortensity perturbation for the turbulent run, at the same time. For the turbulence amplitude taken here, the vortensity perturbations due to turbulence are typically larger than those obtained in the laminar run. Instantaneous streamlines are also depicted as solid curves, which highlight a random walk motion rather than well-defined circulating and librating motions around the (expected) horseshoe region.

However, with turbulence, we are primarily interested in the time-averaged fields. We thus display in the right panel of Figure~\ref{vortplots} the perturbed vortensity time-averaged between 400 and 600 orbits. To calculate this time-average, fields outputs were produced every $1/20^{\rm th}$ of orbit, which is approximately half the lifetime of the turbulent mode with $m=6$. Solid lines depict streamlines averaged over the same time period, using the same time sample. The two streamlines passing near the planet location show the time-averaged separatrices of the mean horseshoe region. All other streamlines of the left and right panels of Figure~\ref{vortplots} are calculated with the same initial coordinates in the $r-\varphi$ plane, and are therefore directly comparable. We comment that the comparison with the instantaneous fields and streamlines of the laminar run is fair, as a steady state is reached well before 400 orbits for this run. In the turbulent run, fluid elements have, in time-average, librating streamlines inside of a mean horseshoe region, and circular streamlines outside of it. The averaged streamlines of the turbulent run much resemble the instantaneous streamlines of the laminar run. In particular, the half-width of the mean horseshoe region with turbulence does not significantly differ from that without turbulence. The time-averaged vortensity perturbations are also very similar to those of the laminar run. This qualitative agreement signifies that both simulations do have similar values of the vortensity's diffusion coefficient, as {\it a priori} expected from our relationship between $\gamma$ and $\langle\alpha_D\rangle$.

The above comparison leads us to the following comments. In time-average, turbulence tends to "diffuse" vortensity, analogous to the effect of viscosity in laminar disks. This similarity could be surprising at first glance since the turbulent potential cannot act as a source term in the vortensity equation. With our turbulent potential, vortensity is therefore conserved along {\it instantaneous} streamlines. Nonetheless, random motions triggered by turbulence cause turbulent diffusion, so that the vortensity along {\it time-averaged} streamlines is not conserved. As shown in \S~\ref{sec:schmidt}, this process can be modeled by a simple diffusion law, featuring the diffusion coefficient $\langle\alpha_D\rangle$. In the appendix, we derive the time-averaged vortensity equation for a two-dimensional disk subject to the turbulent potential. 

\begin{figure}
  \includegraphics[width=\hsize]{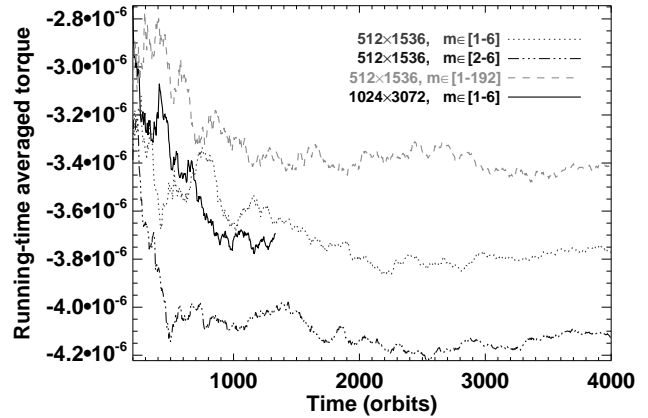}
  \caption{\label{convergence}Convergence properties of the running-time averaged torque for the turbulent runs of \S~\ref{sec:issues}. The torque obtained with our standard resolution ($512\times 1536$) and with the $m>6$ modes cut-off is depicted with a dotted line. The dot-dot-dot-dashed curve displays the torque obtained with the $m<2$ and $m>6$ modes cut-off, at the same resolution. The dashed curve shows the torque without the $m>6$ modes truncation, at the same resolution. The solid curve displays the torque with the $m>6$ modes truncation, at higher resolution ($1024\times 3072$).}
\end{figure}

\subsection{Numerical issues}
\label{sec:issues}
All the calculations with an embedded protoplanet were obtained with a grid resolution of $512\times 1536$ and included the $m>6$ modes truncation in the expression of the turbulent potential. We investigate here how these assumptions affect our results. For this purpose, we consider the turbulent model of \S~\ref{sec:desaturation} with $q = 5\times 10^{-6}$, $\sigma=0.5$ and $\gamma = 4\times 10^{-5}$. Three additional calculations were performed, (i) one at the same resolution but without the $m>6$ modes truncation, (ii) another one at the same resolution with cut-off of both $m=1$ and $m>6$ modes, and (iii) one with the $m>6$ modes truncation but with double resolution ($1024\times 3072$). The running-time averaged torques obtained with these four calculations are displayed in Figure~\ref{convergence}. Doubling the resolution in each direction has no significant impact on the torque evaluation. Although not shown here, we checked that this result holds without the $m>6$ modes truncation. This convergence in resolution is not surprising because our model does not include any feedback from the smallest length scales, where energy dissipates, to the largest ones, through which exchange of angular momentum primarily occurs. The reason for this is that turbulent modes are always regenerated independently of all other modes still at work in the disk. Including an energy equation should not alter the convergence in resolution.

The torque r.t.a. obtained with the additional $m=1$ modes cut-off is more negative, as expected. Discarding $m=1$ modes decreases indeed the turbulence strength, and therefore the equivalent alpha viscosity $\langle\alpha_D\rangle$ associated to turbulence. For our value of $\gamma$, decreasing $\langle\alpha_D\rangle$ weakens the horseshoe drag (see figure~\ref{finalrta}), and the total torque is more negative. The relative difference of the total torques is $\approx 10\%$. From the left panel of figure~\ref{finalrta}, we point out that decreasing $\gamma$ by about one order-of-magnitude would make the torque obtained with the $m>6$ modes truncation almost coincide with the differential Lindblad torque, which is about $50\%$ larger. This comparison underscores that, as anticipated in \S~\ref{sec:rta}, the absence of $m=1$ modes in the study of LSA04 cannot account for the order-of-magnitude difference in the expressions for the turbulent torque r.t.a. given by LSA04 and by Equation~(\ref{sigmaturb}).

\begin{figure}
  \includegraphics[width=\hsize]{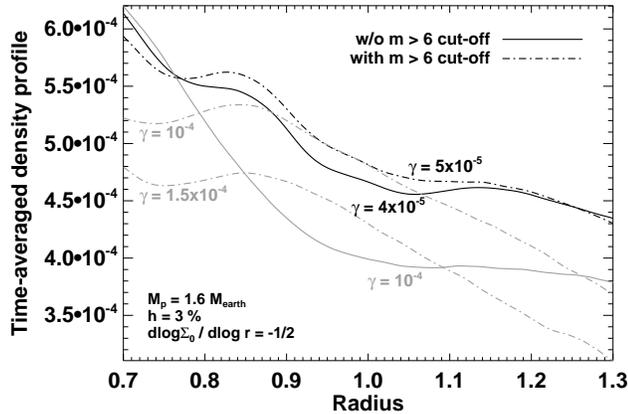}
  \caption{\label{convergence2}Surface density profile time-averaged over $4000$ orbits for turbulent runs with and without the $m>6$ modes cut-off on the expression for the turbulent potential.}
\end{figure}

Discarding the $m>6$ modes cut-off renders the stationary r.t.a. torque more positive. This is also an expected result, since including $m>6$ modes is found to increase $\langle\alpha_R\rangle$ by a factor of $\approx 1.5$, as shown in \S~\ref{sec:alpha}. Assuming that $\langle\alpha_D\rangle$ is increased by the same factor, and because $\gamma \propto \langle\alpha_D\rangle^{1/2}$, we expect that the r.t.a. torques obtained (i) with $\gamma = 4\times 10^{-5}$ and without truncation, and (ii) with $\gamma = 5\times 10^{-5}$ but with truncation, should be approximately the same. For the former run, the steady torque is $\sim -3.4\times 10^{-6}$ (see figure~\ref{convergence}), whereas for the latter run, it amounts to $\sim -3.35\times 10^{-6}$ (see left panel of Figure~\ref{finalrta}). This close agreement confirms that, for $\gamma = 4\times 10^{-5}$, the $m>6$ modes cut-off slightly decreases the turbulence strength. The surface density profiles obtained in previous cases (i) and (ii), and time-averaged over 4000 orbits, are displayed in Figure~\ref{convergence2}. Also, we have investigated the impact of the $m>6$ modes cut-off with $\gamma = 10^{-4}$. The r.t.a. torque over $4000$ orbits reaches a steady value of $\sim -1.5\times 10^{-6}$, which is approximately a factor of $2$ and $2.5$ smaller than the torques with cut-off obtained for $\gamma = 10^{-4}$ and $\gamma = 1.5\times 10^{-4}$, respectively (see left panel of Figure~\ref{finalrta}). Clearly, for $\gamma = 10^{-4}$, the impact of the $m>6$ modes cut-off on the torque cannot be explained by a slight decrease of $\gamma$. Figure~\ref{convergence2} shows that the corresponding density profile (solid gray curve) significantly differs from the density profiles with no cut-off, and obtained for $\gamma = 10^{-4}$ and $\gamma = 1.5\times 10^{-4}$ (dash-dotted gray curves). Interestingly, for this turbulence amplitude, the time-evolution of the disk density profile changes quite substantially without the $m>6$ modes cut-off. Around the planet location, the averaged density profile is flatter without cut-off, which increases the positive horseshoe drag, and it takes smaller values, which also makes the total torque more positive. 

\section{Discussion and conclusions}
\label{sec:discuss}
We have revisited the interaction of a low-mass planet with an isothermal turbulent disk, with a special emphasis on the horseshoe drag desaturation by turbulence. Two-dimensional hydrodynamic simulations were performed, using the turbulence model of \cite{lsa04}. This model is based on applying a turbulent potential to the disk, which corresponds to the superposition of simultaneous wave-like modes.

An in-depth analysis of the turbulent potential properties is undertaken in \S~\ref{sec:turb}. We show that these properties can be selected so that perturbations generated by the turbulent potential much resemble those obtained in 3D MHD calculations. For instance, the modes lifetimes taken by \cite{lsa04} are reduced by a factor of ten to attain an autocorrelation time of order one planet's orbital period. We quantify the transport of angular momentum by measuring Reynolds alpha parameters $\langle\alpha_R\rangle$, which we relate to the turbulence amplitude $\gamma$ in Equation~(\ref{alpha}). The effect of turbulence diffusion is also quantified by evaluating vortensity's diffusion coefficients, which we model by an equivalent alpha viscosity $\langle\alpha_D\rangle$. A simple relation between $\gamma$ and $\langle\alpha_D\rangle$ is given by Equation~(\ref{alpha2}). An important result is that vortensity diffusion is about four times more efficient than transport of angular momentum: $\langle\alpha_D\rangle \approx 4\langle\alpha_R\rangle$. We argue in \S~\ref{sec:schmidt} that $\langle\alpha_D\rangle$ can therefore model the total alpha parameter of typical 3D MHD calculations (assuming they have a radial Schmidt number of order unity). Should our results be checked against such calculations, one would have to set up the initial conditions such that the average Reynolds alpha parameter $\langle\alpha_R\rangle$ be similar to one of our values (but note that torque differences will naturally arise from the transition from 2D to 3D). We also provide in Equation~(\ref{sigmaturb}) an expression for the mean deviation of the turbulent torque distribution that gives good agreement with 3D MHD simulations of a disk fully invaded by the magnetorotational instability. Our turbulence model should therefore be well-suited for studying disk-planet interactions not only in dead zones, where angular momentum transport is mostly transported by density waves \citep{flemingstone03, oishi07}, but also in active layers, where turbulence primarily originates. Note however that our turbulence model has compressible modes only (turbulence is driven by a scalar potential), whereas MRI turbulence also contains incompressible (vortical) modes.

By using accurate estimates of the running-time averaged turbulent torque, and of the fully unsaturated torque expected in laminar disks, we give in Equation~(\ref{tauconv3}) an upper estimate (see \S~\ref{sec:convtime}) of the convergence time in our simulations. This convergence time, which is the timescale such that the running-time averaged stochastic torque becomes a small fraction of the total laminar torque, is proportional to $M^{-1/2}_p h_p^2$. It is thus particularly sensitive to the disk scale height. For our conservative disk and planet parameters ($M_p \approx 1.6M_{\oplus}$, $h_p = 3\%$, $\sigma=0.5$, $\tau=0$), up to $\sim 2000$ planet orbital periods are required before the running-time averaged turbulent torque attains $\sim 10\%$ of the fully unsaturated torque expected in a similar laminar model. Increasing the planet mass to $M_p = 10M_{\oplus}$, and the disk aspect ratio to $h_p = 7\%$, yields a maximum convergence time of $\sim 4500\,T_{\rm orb}$.

We then present in \S~\ref{sec:numres} our calculation results with a low-mass planet embedded in a turbulent disk. For comparison, similar laminar calculations are performed with a kinematic viscosity $\nu = \alpha c_s H$. The main results are the following:
\begin{itemize}
\item The averaged differential Lindblad torque with turbulence takes very similar values than in laminar disk models, providing (i) the former is time-averaged over a sufficiently long time period, and (ii) turbulence does not significantly alter the time-averaged density profile.
\item Turbulence can unsaturate the horseshoe drag, depending on the turbulence strength. The horseshoe drag desaturation by turbulence can be modeled by vortensity diffusion across the time-averaged horseshoe region, with a diffusion coefficient $D = \langle\alpha_D\rangle c_s H$. We comment that it is unclear how the horseshoe drag behaves when the largest size of the turbulence eddies becomes comparable to, or larger than the width of the horseshoe region. Vortensity should then enter the mean horseshoe region in an advective way, rather than in a diffusive way. This situation deserves attention, particularly for low-mass planets. It is possible that the horseshoe drag value is determined by the advection timescale across the mean horseshoe region, at the averaged turbulent velocity, in comparison with the libration and U-turn timescales.
\item For similar vortensity's diffusion coefficients ($\alpha \approx \langle\alpha_D\rangle$), time-averaged total torques with turbulence are compared with steady laminar torques in \S~\ref{sec:complam} (see Figure~\ref{finalrta}). At low-turbulence, turbulent and laminar torques are in very good agreement. At high-turbulence, differences arise, which can be fully accounted for by the time-evolution of the averaged density profile with turbulence. These results have two implications. On the one hand, the torque felt by a low-mass planet can be decomposed into a laminar torque, and a stochastic torque due to turbulence. On the other hand, the averaged value of the horseshoe drag in turbulent and laminar disk models are very similar. The same comment applies to the structure of the mean horseshoe region, as shown in \S~\ref{sec:hsregion}. It indicates that the criterion for the horseshoe drag desaturation, $\tau_{\rm u-turn} < \tau_{\rm visc} < \tau_{\rm lib}$, still holds in turbulent disks, in time-average, at least for the turbulence amplitudes considered in our study (but see the above comment on the advective regime). The quantity $\tau_{\rm visc}$ corresponds to $\tau_{\rm visc} \approx x_s^2 / D$, where $D = \langle\alpha_D\rangle c_s H$ is the vortensity's averaged diffusion coefficient, and $x_s$ is the half-width of the mean horseshoe region.
\end{itemize}

With turbulence properties as close as possible to those of 3D MHD simulations, we find that the horseshoe drag exerted by isothermal disks on low-mass planets can remain unsaturated on the long term, depending on the turbulence strength. These results require confirmation by 3D MHD long-term simulations of planet-disk interactions, with disks either fully magnetized or harboring a dead zone. It would also be of relevant interest to investigate how accretion onto a low-mass planet can be affected by turbulent motions in the planet vicinity. In forthcoming works, we will extend our study to radiative disks, and we will investigate the trapping of a protoplanet at a density transition in presence of turbulence.

\acknowledgements
It is a pleasure to thank S{\'e}bastien Fromang, Fr{\'e}d{\'e}ric Masset and Richard Nelson for illuminating conversations and very fruitful suggestions. The authors are thankful to Fr{\'e}d{\'e}ric Masset for a thorough reading of a first draft of this manuscript, and to the anonymous referee for an insightful report. We also thank Nic Brummell, Greg Laughlin, and John Papaloizou for useful discussions. C.B. is grateful to the Kavli Institute for Astronomy and Astrophysics for its kind hospitality and support during a portion of this work. Computations were performed on the Pleiades Cluster at UC Santa Cruz. This work is supported by NASA (NNG06-GF45G, NNX07A-L13G, NNX07AI88G) and NSF(AST-0507424).

\appendix
\section{Time-averaged vortensity equation}
We derive in this section the time-averaged vortensity equation for a two-dimensional disk subject to the turbulent potential described in \S~\ref{sec:turb}. In a frame rotating uniformly at angular velocity $\Omega$, the continuity and momentum equations are
\begin{equation}
\frac{\partial \Sigma}{\partial t} + {\bf v} \cdot {\bf \nabla} \Sigma + \Sigma {\bf \nabla} \cdot {\bf v} = 0,
\label{app:mass}
\end{equation}
\begin{equation}
\frac{\partial {\bf v}}{\partial t}  +  ({\bf v} \cdot {\bf \nabla}){\bf v}  + 2\Omega {\bf k} \times {\bf v} = -{\bf \nabla} (h + \Phi),
\label{app:mom1}
\end{equation}
where ${\bf v}$ is the two-dimensional velocity field. In Equation~(\ref{app:mom1}), $h = \int dp/\Sigma$ is the fluid enthalpy, $\Phi$ is the total potential felt by the disk (including the time-dependent turbulent potential), and ${\bf k}$ is the unit vector in the vertical direction. Using the vector identity $({\bf v} \cdot {\bf \nabla}) {\bf v} = (1/2) {\bf \nabla} {\bf v}^2 - {\bf v} \times {\bf w}$, where ${\bf w}= {\bf \nabla} \times  {\bf v}$ is the vorticity, and taking the curl of Equation~(\ref{app:mom1}), we find
\begin{equation}
\frac{\partial ({\bf w} + 2 \Omega {\bf k})}{\partial t} +  {\bf \nabla} \times [( {\bf w}+2 \Omega {\bf k}) \times {\bf v}] = {\bf 0}.
\label{app:mom}
\end{equation}
Multiplying Equation~(\ref{app:mom}) by $\Sigma^{-1}$, and subtracting the product of $({\bf w} + 2 \Omega {\bf k}) / \Sigma^2$ and Equation~(\ref{app:mass}), we are left with
\begin{equation}
\frac{\partial {\bf \xi}}{\partial t} + ({\bf v} \cdot {\bf \nabla}){\bf \xi} =  {\bf 0},
\label{app:xi}
\end{equation}
where ${\bf \xi} = ({\bf w} + 2\Omega {\bf k})/\Sigma$ is the vortensity. In the above derivation,  we used the vector identity ${\bf \nabla} \times ( {\bf A} \times {\bf B}) = {\bf A} \cdot ({\bf \nabla} \cdot {\bf B}) - ({\bf A} \cdot {\bf \nabla}) {\bf B} - {\bf B} \cdot ({\bf \nabla} \cdot {\bf A}) + ({\bf B} \cdot {\bf \nabla}) {\bf A}$. Equation~(\ref{app:xi}) reduces to equation 2 in \cite{korpap96} in the case of a steady flow. We use the so-called Reynolds decomposition, in which an instantaneous value is written as the sum of a mean value (denoted with a zero subscript) plus a fluctuation: ${\bf v} = {\bf v_0} + {\bf \delta v}$ and ${\bf \xi} = {\bf \xi_0} + {\bf \delta\xi}$. Mean values are taken over a timescale $T$ that is large compared to the correlation timescale, but short compared to that of the flow evolution. We thus have $\overline{{\bf v}} = T^{-1} \int_0^T {\bf v} dt = {\bf v_0}$ and $\overline{{\bf \delta v}}= {\bf 0}$. Similarly, $\overline{{\bf \xi}} = {\bf \xi_0}$ and $\overline{{\bf \delta\xi}} = {\bf 0}$. After time-averaging Equation~(\ref{app:xi}), we find
\begin{equation}
\frac{\partial \overline{{\bf \xi}}}{\partial \tilde{t}}+ 
(\overline{{\bf v}} \cdot {\bf \nabla}) \overline{{\bf \xi}} + 
\overline{({\bf \delta v} \cdot {\bf \nabla}) {\bf \delta \xi}} = {\bf 0},
\end{equation}
where the time interval between two successive values of $\tilde{t}$ is greater or equal than $T$, and where the term $\overline{({\bf \delta v} \cdot {\bf \nabla}) {\bf \delta \xi}}$, which features the correlation between ${\bf \delta v}$ and ${\bf \delta \xi}$, is responsible for the non-conservation of vortensity along time-averaged streamlines. It is analogous to the Reynolds stress in the averaged momentum equation.


\end{document}